\documentclass[a4paper,11pt]{article}
\pdfoutput=1

\usepackage[utf8]{inputenc}
\usepackage{a4wide}
\usepackage{graphicx}
\graphicspath{{figures/}}
\usepackage[small,bf]{caption}
\usepackage{amssymb}
\usepackage{amsmath}
\usepackage{slashed}
\usepackage{mathtools}
\usepackage{cite} 
\usepackage[compat=1.1.0]{tikz-feynman}

\newcommand{\be}{\begin{equation}}
\newcommand{\ee}{\end{equation}}
\newcommand{\ben}{\begin{eqnarray}}
\newcommand{\een}{\end{eqnarray}}
\newcommand{\bi}{\begin{itemize}}
\newcommand{\ei}{\end{itemize}}

\DeclarePairedDelimiter{\evdel}{\langle}{\rangle}
\newcommand{\ev}{\evdel} 

\DeclareMathOperator{\diag}{diag}
\DeclareMathOperator{\sign}{sgn}

\def\nn{\nonumber}
\def\d{\partial}
\def\a{\alpha}

\def\g{\gamma}

\def\l{\lambda}

\def\s{\sigma}

\def\th{\theta}

\def\O{\Omega}

\def\vr{v_r}

\def\eps{\epsilon}

\numberwithin{equation}{section}
\makeatletter
\g@addto@macro\bfseries{\boldmath}
\makeatother

\begin{document}

\begin{titlepage}
\begin{flushright}
FTUAM-19-5\\
IFT-UAM/CSIC-19-19\\
IPPP/19/17 \\
\end{flushright}
\vspace*{0.8cm}

\begin{center}
{\Large\bf Neutrino Portals to Dark Matter}\\[0.8cm]
M.~Blennow,$^{a,b}$
E.~Fernandez-Martinez,$^a$
A.~Olivares-Del Campo,$^c$\\
S.~Pascoli,$^c$
S.~Rosauro-Alcaraz,$^a$
and A.~V.~Titov$^{\,c}$\\[0.4cm]
$^{a}$\,{\it Departamento de F\'isica Te\'orica and Instituto de F\'{\i}sica Te\'orica, IFT-UAM/CSIC,\\
Universidad Aut\'onoma de Madrid, Cantoblanco, 28049, Madrid, Spain} \\
$^{b}$\,{\it Department of Physics, School of Engineering Sciences, \\ KTH Royal Institute of Technology, AlbaNova University Center, \\ Roslagstullsbacken 21, SE--106 91 Stockholm, Sweden}\\
$^{c}$\,{\it Institute for Particle Physics Phenomenology, 
Department of Physics, Durham University,\\ 
South Road, Durham DH1 3LE, United Kingdom}
\end{center}
\vspace{0.8cm}

\begin{abstract}
We explore the possibility that dark matter interactions with Standard Model particles are dominated by interactions with neutrinos. We examine whether it is possible to construct such a scenario in a gauge invariant manner. We first study the coupling of dark matter to the full lepton doublet and confirm that this generally leads to the dark matter phenomenology being dominated by interactions with charged leptons. We then explore two different implementations of the neutrino portal in which neutrinos mix with a Standard Model singlet fermion that interacts directly with dark matter through either a scalar or vector mediator. In the latter cases we find that the neutrino interactions can dominate the dark matter phenomenology. Present neutrino detectors can probe dark matter annihilations into neutrinos and already set the strongest constraints on these realisations. Future experiments such as Hyper-Kamiokande, MEMPHYS, DUNE, or DARWIN could allow to probe dark matter-neutrino cross sections down to the value required to obtain the correct thermal relic abundance.
\end{abstract}
\end{titlepage}

\section{Introduction}
The unknown origin of neutrino masses and mixing together with the existence of the dark matter (DM) component of the Universe constitute our most significant experimental evidence for physics beyond the Standard Model (SM) and therefore the best windows to explore new physics. Neutrinos and DM also share an elusive nature with very weak interactions with the other SM particles. Indeed, neutrinos only participate in the weak interactions of the SM while all direct and indirect searches for DM interactions with the SM, other than gravity, are so far negative or inconclusive. A tantalising avenue of investigation is the possibility of a stronger connection between these two sectors. In this case, the best way to probe DM would be through the neutrino sector.

Several works have investigated the phenomenology of a dominant interaction between the neutrino and DM sectors and the possibility to probe DM through neutrinos both via its cosmological implications~\cite{Boehm:2000gq,Boehm:2004th,Mangano:2006mp,Bertschinger:2006nq,Serra:2009uu,Aarssen:2012fx,Bringmann:2013vra,Wilkinson:2014ksa,Boehm:2014vja,Schewtschenko:2014fca,Cyr-Racine:2015ihg,Vogelsberger:2015gpr,Balducci:2017vwg,Campo:2017nwh} as well as through indirect searches~\cite{Beacom:2006tt,PalomaresRuiz:2007eu,Lindner:2010rr,ElAisati:2017ppn,Campo:2017nwh}. In the presence of this interaction, DM would no longer be collisionless, but able to scatter with neutrinos in the Early Universe, affecting matter density fluctuations. 
Moreover, the power spectrum would show a suppression at small scales~\cite{Boehm:2014vja,Schewtschenko:2014fca,Campo:2017nwh} or even an oscillatory pattern~\cite{Bertschinger:2006nq,Mangano:2006mp,Serra:2009uu,Wilkinson:2014ksa}.
Indirect detection searches for DM annihilating to neutrinos in the galactic centre
have also been performed at neutrino detectors and used to constrain DM-neutrino interactions~\cite{Beacom:2006tt,PalomaresRuiz:2007eu,Campo:2017nwh}.
The propagation of neutrinos through DM halos could be modified as well, leading to dips in supernova neutrino spectra due to resonant interactions with DM~\cite{Farzan:2014gza,Franarin:2018gfk}, or affect the spectrum or isotropy of the high energy cosmic neutrinos observed by IceCube~\cite{Arguelles:2017atb,Pandey:2018wvh,Karmakar:2018fno}.  

However, it is not straightforward to envision a scenario in which the neutrino-DM interactions dominate the DM phenomenology. Naively, gauge invariance dictates that the interactions of the left-handed (LH) SM neutrinos with DM will be equal to those of their charged lepton counterparts in the $SU(2)$ doublets. In this case, the best windows to DM would instead be the charged leptons rather than the more elusive neutrinos.

In this work, we will investigate some gauge-invariant SM extensions that lead to sizeable neutrino-DM interactions,
exploring if neutrino probes could dominate our sensitivity to the dark sector. This is actually a rather natural possibility. In fact, if DM does not participate in any of the SM gauge interactions, 
the natural expectation is
that the strongest connection to DM will be via singlets of the SM gauge group. Indeed, if non-singlet fields were involved instead, the dimensionality of the operators linking the two sectors would have to increase in order to comply with gauge invariance. This reasoning leads to the three well-known SM portals to the dark sector: the ``gauge boson portal''~\cite{Holdom:1985ag}, the ``Higgs portal''~\cite{Patt:2006fw}, and the ``neutrino portal''~\cite{Falkowski:2009yz,Lindner:2010rr,Macias:2015cna}. The neutrino portal includes the addition of right-handed (RH) neutrinos $N_R$, which makes this option particularly appealing in connection to the evidence of neutrino masses and mixing from neutrino oscillations.

Since the neutrino portal relies on the mixing between $N_R$ and the light SM neutrinos to connect the neutrino and DM sectors, this mixing needs to be sizeable. In the ``canonical'' seesaw mechanism~\cite{Minkowski:1977sc,Yanagida:1979as,GellMann:1980vs,Glashow:1979nm,Mohapatra:1979ia}, the smallness of neutrino masses is explained through a large Majorana mass for $N_R$ and the mixings are then similarly suppressed by the large scale. This option, which is rather natural from the point of view of neutrino masses, worsens the Higgs hierarchy problem~\cite{Casas:2004gh}. An interesting alternative is to explain the smallness of neutrino masses via a symmetry argument instead~\cite{Mohapatra:1986bd,Bernabeu:1987gr,Branco:1988ex,Buchmuller:1990du,Pilaftsis:1991ug,Kersten:2007vk}. Indeed, in models with an approximate lepton number ($L$) symmetry such as the linear~\cite{Malinsky:2005bi} or inverse~\cite{Mohapatra:1986bd,Bernabeu:1987gr} seesaw mechanisms, neutrino masses are suppressed by the small $L$-breaking parameters while light neutrino mixing with $N_R$ is unsuppressed. 
In the present study, we will assume relatively large mixing angles noting that they can be compatible with neutrino masses, but we will not specify a concrete neutrino mass generation mechanism, since these small lepton number violating parameters, and hence light neutrino masses, will have no significant impact on the DM-related phenomenology.

We will consider fermionic DM and, more specifically, Dirac DM, which has the richest phenomenology when interacting with SM neutrinos. Indeed, the dominant term in the annihilation cross section to neutrinos is not velocity suppressed, and DM annihilations therefore lead to interesting signatures in indirect searches. Alternative scenarios with a Majorana, scalar, or vector DM candidate will lead to a velocity-dependent annihilation cross section to neutrinos~\cite{Campo:2017nwh}. While such possibilities are viable, they are difficult to probe experimentally at neutrino detectors. This is due to the fact that the DM velocity in the halo today is $v_{\rm{halo}} = 10^{-3}c$~\cite{Bertone:2010zza}, which significantly reduces the annihilation rate to neutrinos.

The article is organised as follows. 
In Section~\ref{sec:constraints}, we summarise relevant experimental searches for DM 
and constraints coming from cosmology. 
In Section~\ref{sec:DMEFT}, we consider the simplest 
gauge-invariant scenario, in which DM is coupled directly 
to the full SM lepton doublet. 
In this case, as expected, the charged lepton probes tend to dominate 
the constraints on the DM parameter space.
Further, in Section~\ref{sec:NPortal},
we introduce the neutrino portal involving 
one new Dirac sterile neutrino $N$,
which will communicate with the dark sector. We present two realisations of the neutrino portal, for scalar~\cite{Falkowski:2011xh,Bertoni:2014mva,Gonzalez-Macias:2016vxy,Batell:2017cmf} and
vector~\cite{Cherry:2014xra} interactions between the DM and $N$ in Sections~\ref{sec:Scalar} and \ref{sec:Vector}, respectively. 
For both of them, we investigate the parameter space, 
demonstrating that current and future neutrino experiments have 
the dominant role in constraining it. 
Finally, we conclude in Section~\ref{sec:conclusions}.

\section{Constraints on interactions of DM with SM particles}
\label{sec:constraints}
In the next sections, we will explore the parameter space of different possible gauge-invariant ways to realise interactions of neutrinos with DM. For each realisation, we will investigate whether it is possible for these DM-neutrino interactions to play a dominant role in the DM phenomenology. In particular, we will address whether or not the DM relic abundance can be achieved via the DM-neutrino interactions and/or if indirect DM searches via its annihilation into neutrinos (probed at neutrino detectors) can be the dominant test of the model parameter space. We will use the observables presented in this section to place constraints on the parameter space of each scenario.

\subsection{Indirect detection searches for DM annihilation to neutrinos}
\label{IDSN}
DM annihilating in high density regions such as the Milky Way can generate a significant monochromatic flux of neutrinos with energy $E_\nu = m_\chi$, where $m_\chi$ is the DM mass. This flux is proportional to the integral of the DM density squared along the line of sight and can be searched for in neutrino detectors such as Super-Kamiokande (SK)~\cite{Fukuda:2002uc} or Borexino~\cite{Alimonti:2000xc}. 

Several analyses that use neutrino detectors to probe the DM parameter space have been performed in the literature~\cite{Beacom:2006tt,PalomaresRuiz:2007eu,Yuksel:2007ac,Frankiewicz:2015zma,ElAisati:2017ppn,Campo:2017nwh,Campo:2018dfh,Klop:2018ltd}. For small DM masses in the range $2 - 17$~MeV, 
we can exploit the upper bound on the monochromatic antineutrino flux 
set by Borexino~\cite{Bellini:2010gn} and convert it to a conservative upper bound of 
$\ev{\s\vr} \lesssim 10^{-22} - 10^{-20}$~cm$^3/$s on the thermally averaged annihilation cross section $\sigma$ multiplied by the relative velocity $\vr$ of DM particles, as discussed in Ref.~\cite{Campo:2017nwh}. Likewise, between $10$ and $200$~MeV, SK can place an upper bound of $\ev{\s\vr} \lesssim 10^{-25} - 10^{-23}$~cm$^3/$s (depending on the DM mass)~\cite{Campo:2017nwh}. For DM with a mass between $1$~GeV and $10$~TeV annihilating in the galactic centre, 
the SK collaboration has performed a dedicated  analysis
and set an upper bound of $\ev{\s\vr} \sim 10^{-24} - 10^{-22}$~cm$^3/$s \cite{Frankiewicz:2015zma}. We will also consider the general upper bound on $\ev{\s\vr}$
derived in Ref.~\cite{Beacom:2006tt} by calculating the cosmic diffuse neutrino signal 
from DM annihilations in all halos in the Universe
and comparing it to the measured 
atmospheric neutrino background
by Fr\'ejus~\cite{Berger:1987ke}, AMANDA~\cite{Halzen:1998bp}, and SK. 
This bound applies to $m_\chi$ in the range between $100$~MeV and $100$~TeV 
and excludes $\ev{\s\vr} \gtrsim 10^{-23} - 10^{-21}$~cm$^3/$s (depending on $m_\chi$). 
As argued in Ref.~\cite{Beacom:2006tt}, this bound could be improved
by one or even two orders of magnitude with dedicated analyses 
by existing neutrino experiments such as SK. 

The next generation experiment Hyper-Kamiokande (HK)~\cite{Abe:2018uyc} will be sensitive to approximately one order of magnitude smaller cross sections in this mass range. 
Indeed, with a 187~kton fiducial mass and an exposure time of 10~years, HK could probe the parameter space almost down to the relic density cross section ($\ev{\s\vr} = 3\times10^{-26}$~cm$^3/$s \cite{Jungman:1995df}). Possible improvements such as additional mass from a second tank together with Gd doping for background reduction would allow to probe beyond this value~\cite{Campo:2018dfh}. Similarly, the ESS$\nu$SB project~\cite{Baussan:2013zcy} envisions a 500~kton fiducial water detector, MEMPHYS~\cite{deBellefon:2006vq}, that would have slightly better sensitivity than HK from the additional fiducial mass. Similarly, future DM and neutrino detectors such as DARWIN~\cite{Aalbers:2016jon} and DUNE~\cite{Acciarri:2015uup} will be able to further constrain the DM annihilation cross section to neutrinos. 
DARWIN will set stronger bounds for DM masses between $100$~MeV and $1$~GeV~\cite{McKeen:2018pbb}, while DUNE will be able to exclude thermal DM masses between $25$ and $100$~MeV~\cite{Klop:2018ltd}. 

Competitive constraints from DM annihilations in the Sun to neutrinos, or other SM particles that decay to neutrinos, have also been derived by neutrino detectors such as SK~\cite{Choi:2015ara} and IceCube~\cite{Aartsen:2016zhm}. These exploit the higher DM concentration expected in the solar interior since it could capture DM particles from the halo via scatterings. In all the realisations under study we explore the connection between the DM and neutrino sectors with very suppressed interactions with the rest of the SM, in particular with quarks. Thus, in these scenarios, the Sun does not accrete DM particles effectively and the constraints from these searches do not apply.

\subsection{Indirect detection searches for DM annihilation to charged leptons}
\label{IDSCL}
DM interactions with charged leptons will always be present either at tree level, if DM couples to the full doublet, 
or at loop level in the neutrino portal scenarios. 
Therefore, we will take into account
indirect detection searches for DM annihilations to charged leptons from the Fermi satellite~\cite{Ahnen:2016qkx}, 
as well as from their 
imprint in the cosmic microwave background (CMB) 
as observed by  Planck~\cite{Aghanim:2018eyx,Slatyer:2015jla}.

\subsection{Direct detection searches}
\label{DDS}
DM will not couple directly to the quarks in any of the scenarios that we will discuss. Nevertheless, such couplings will arise at loop level 
in a similar way to the DM-charged lepton interactions. As we will see, bounds from direct detection experiments, such as XENON1T~\cite{Aprile:2018dbl}, are so stringent that they will still constrain the parameter space for large DM masses. 
Recently, direct detection of sub-GeV DM via scattering off electrons has gained significant attention~\cite{Battaglieri:2017aum,Essig:2017kqs,Essig:2018tss,Dolan:2017xbu}. We have also considered this process and found it to be sub-leading with respect to other relevant constraints.

\subsection{Constraints from cosmology}
\label{CC}
If DM remains in thermal equilibrium with neutrinos during Big Bang nucleosynthesis (BBN), it can spoil its predictions~\cite{Serpico:2004nm,Iocco:2008va}. Similarly, the effective number of neutrinos, as constrained by CMB measurements, would be affected if DM remained in equilibrium after neutrinos decoupled from the photon plasma~\cite{Boehm:2013jpa,Nollett:2014lwa,Escudero:2018mvt}. Thus, to avoid these two effects, we will not consider DM masses $m_\chi < 10$~MeV. Moreover, DM-neutrino interactions can also have an effect in the formation of large scale structures (LSS) since, as DM particles scatter off neutrinos, they diffuse out and erase small scale perturbations. This effect leads to a suppression of the amount of small scale structures today. By comparing LSS predictions to observations, one can set an upper bound on the strength of the elastic scattering between DM and neutrinos~\cite{Wilkinson:2014ksa,Escudero:2018thh}. Nevertheless, for the models we are presenting in this work, the mixing between the sterile and SM neutrino suppresses the neutrino-DM elastic scattering and, consequently, its effect on LSS constrains regions of the parameter space already ruled out by CMB and BBN constraints~\cite{Campo:2017nwh}.

\section{Coupling to the full lepton doublet}
\label{sec:DMEFT}
In this section, we will study the simplest scenario, in which the neutrino-DM interaction arises from a direct coupling to the full SM $SU(2)$ lepton doublet. In order to avoid specifying the nature of the mediator, we will adopt an effective field theory approach, simply adding a $d=6$, 4-fermion interaction.

\subsection{Model}
Since the 4-fermion operator needs to involve two LH SM lepton doublets $L_\a = (\nu_{\a L}, \ell_{\a L})^T$, $\a = e,\mu,\tau$, its Lorentz structure is fixed to be $\overline{L_{\alpha}}\gamma^{\mu}L_{\alpha}$. For definiteness we will assume a vector structure for the DM part. An axial coupling would instead lead to a velocity-suppressed DM annihilation
cross section to neutrinos for both DM relic abundance and indirect searches. The cross section for DM annihilation to charged leptons would however have an additional term only suppressed by the lepton mass, and thus, it would tend to dominate over the annihilation cross section 
to neutrinos. Therefore, we will not consider this option in what follows.

 The Lagrangian describing the neutrino-DM interaction is thus given by
\begin{align}
\mathcal{L} & = \mathcal{L}_{\mathrm{SM}} 
+ \overline{\chi}\left(i\slashed{\d}-m_{\chi}\right)\chi
+\frac{c_{\alpha}}{\Lambda^2}\,
\overline{\chi}\gamma_{\mu}\chi\,
\overline{L_{\alpha}}\gamma^{\mu}L_{\alpha}\,,
\label{eq:NBFPortalLagrangian}
\end{align}
where $\chi$ is a Dirac fermion DM particle, and flavour diagonal couplings $c_\alpha/\Lambda^2$ between DM and the lepton doublets have been assumed in order to avoid new sources of flavour violation. For the effective description to be consistent we will require that $\Lambda^2/c_\alpha \gg m_{\chi}^2$.
The simplest UV completion which leads to the $d=6$ operator in Eq.~\eqref{eq:NBFPortalLagrangian} is via the exchange of a new heavy vector boson that couples both to $\chi$ and $L_\alpha$.

The Lagrangian in Eq.~\eqref{eq:NBFPortalLagrangian} implies that, in this naive gauge-invariant scenario, the coupling between the SM neutrinos and DM will be accompanied by a DM-charged lepton coupling of the same strength. Therefore, the strongest constraints on this model will typically come from indirect searches for DM annihilations to charged leptons. The DM relic abundance will also be set by its annihilation into leptons, either neutrinos or charged leptons, with the annihilation cross section given by
\be
\langle \sigma v_r \rangle \approx \frac{c_{\alpha}^2 m_\chi^2}{2\pi\Lambda^4}
\left(1-\frac{m_{\alpha}^2}{4m_\chi^2}\right)
\sqrt{1 - \frac{m_\alpha^2}{m_\chi^2}}\,,
\label{eq:XSBFPortal}
\ee
where $m_{\alpha}$ is the lepton mass for the different $\alpha$ flavour.

\subsection{Results}
\begin{figure}
\centering
\includegraphics[width=7.6cm]{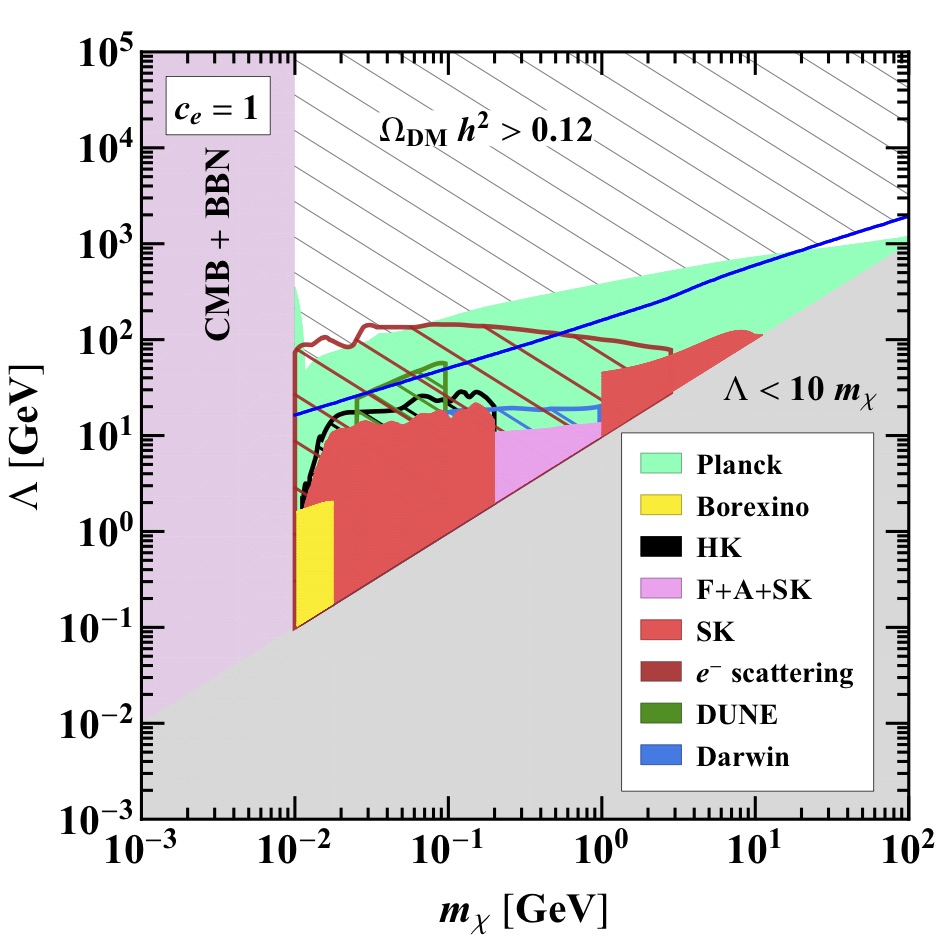}
\hspace{0.1cm}
\includegraphics[width=7.6cm]{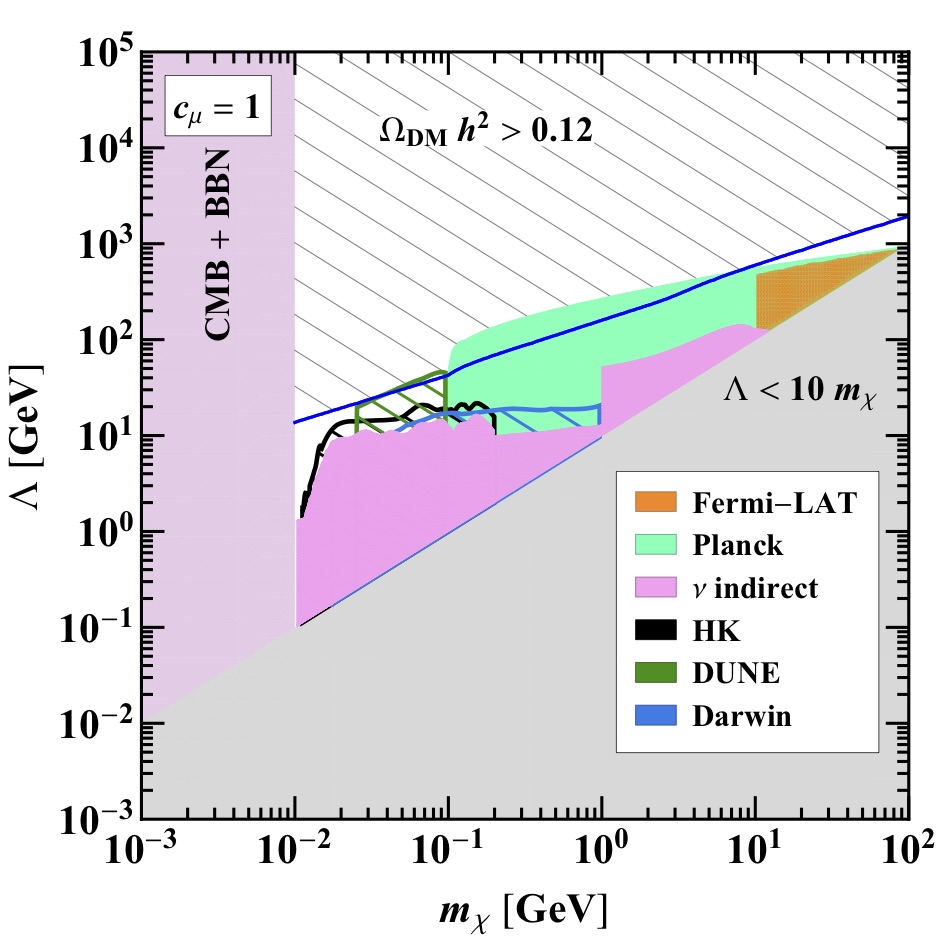}
\includegraphics[width=7.6cm]{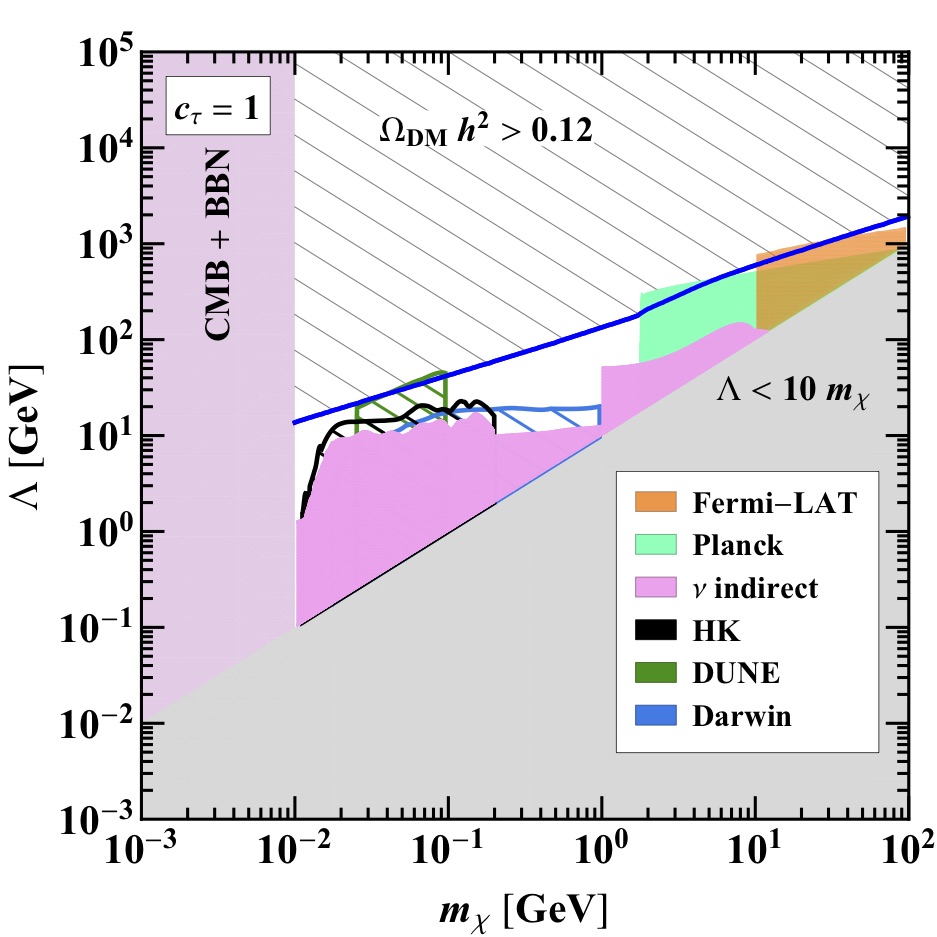}
\hspace{0.1cm}
\includegraphics[width=7.6cm]{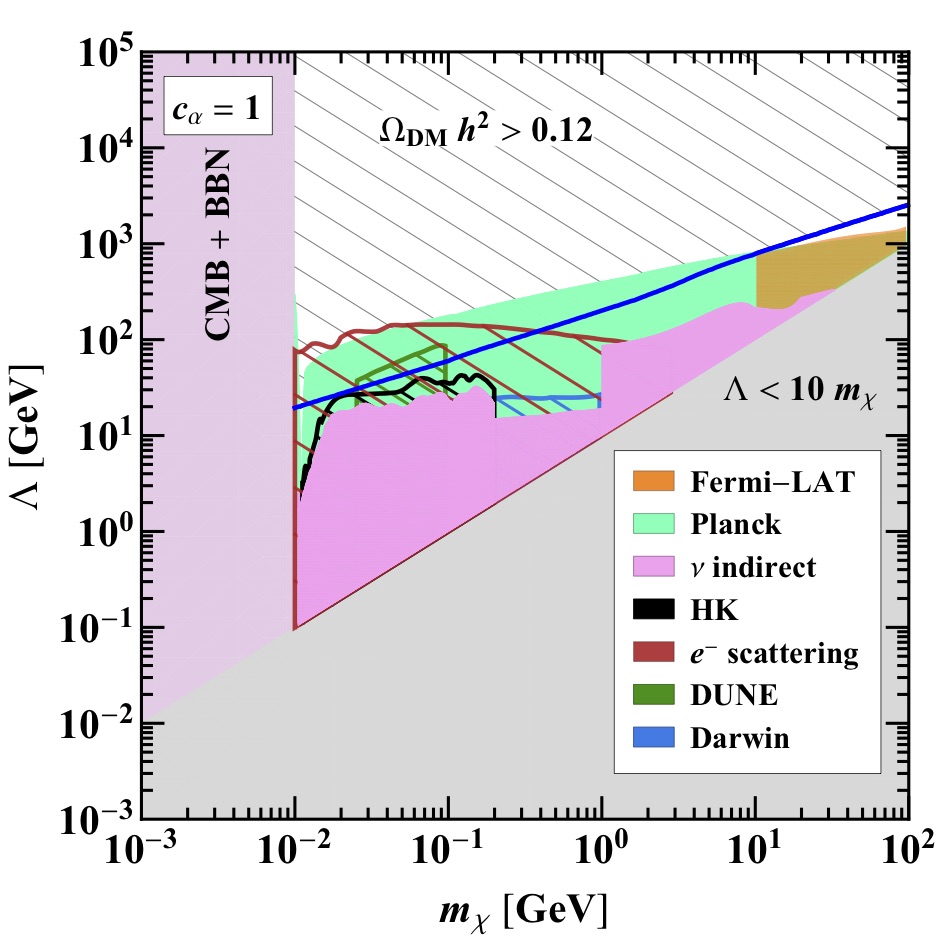}
\caption{Constraints on the DM mass $m_{\chi}$ and the new physics scale $\Lambda$. 
The upper and bottom-left panels correspond to couplings to only one of the lepton doublets (electron, muon, or tau), while the bottom-right panel corresponds to all three couplings being of equal strength.
Along the blue line we recover the correct DM relic abundance from thermal freeze-out. 
The coloured shaded regions are excluded by different experiments, while the hatched areas correspond to prospective sensitivities of future experiments.
The lower bound $m_\chi \gtrsim 10$~MeV 
is set by observations of the CMB and BBN. See text for further details.}
\label{Fig:BruteForceResults}
\end{figure}
%
In Fig.~\ref{Fig:BruteForceResults}, we show regions in the parameter space of the DM mass $m_\chi$ and 
the new physics scale $\Lambda$ excluded by 
different experiments. 
The blue line 
corresponds to 
the correct DM relic density 
$\O_\mathrm{DM} h^2 = 0.1193 \pm 0.0009$~\cite{Aghanim:2018eyx} 
obtained through the thermal freeze-out 
mechanism. This line has been 
computed with \texttt{micrOMEGAs}~\cite{Belanger:2001fz}.
In the upper hatched region, the DM-lepton interaction would be too weak,
leading to overclosure of the Universe ($\O_\mathrm{DM} h^2>0.12$). 
In the region below the blue line, 
the relic density is smaller than 
the observed DM abundance.
If there are additional production mechanisms contributing to the DM density, 
this region is also viable. 

The constraints from indirect DM searches outlined in 
Section~\ref{sec:constraints} are shown as different shaded regions. 
The light green (Planck~\cite{Aghanim:2018eyx, Slatyer:2015jla}) and orange (Fermi satellite~\cite{Ahnen:2016qkx}) regions correspond to the bounds from DM annihilation to charged leptons described in Section~\ref{IDSCL}. The remaining shaded regions correspond to the constraints from DM annihilation to neutrinos 
as searched for in neutrino detectors and summarised in Section~\ref{IDSN}. In the upper-left panel of Fig.~\ref{Fig:BruteForceResults}, we show in different colours the bounds coming from different neutrino experiments. The SK analyses~\cite{Campo:2017nwh, Frankiewicz:2015zma} are shown in red while the Borexino bounds~\cite{Bellini:2010gn} are displayed in yellow.
The pink colour corresponds to
the bounds from~\cite{Yuksel:2007ac} 
obtained by combining the atmospheric neutrino data.%
\footnote{``F+A+SK" in the corresponding legend stands for 
Fr\'ejus + AMANDA + SK.}
The dark red hatched region corresponds to 
prospective sensitivity of experiments on DM-electron scattering~\cite{Essig:2018tss}, while the blue, black, and green hatched regions correspond to prospects from different neutrino experiments as described in Section~\ref{IDSN}. 
In the following panels and in the rest of the paper 
we show all present indirect detection 
constraints from neutrino experiments in pink colour.

As can be seen in Fig.~\ref{Fig:BruteForceResults}, the strongest constraints come from DM annihilation to charged leptons as probed by Fermi-LAT~\cite{Ahnen:2016qkx} for $\chi\overline{\chi} \to \tau^+\tau^-$, $\mu^+\mu^-$ and from Planck~\cite{Aghanim:2018eyx,Slatyer:2015jla} for $\chi\overline{\chi} \to \ell^+ \ell^-$, $\ell = e,\mu,\tau$. 
The latter are in agreement with the 
results of Ref.~\cite{Bertuzzo:2017lwt}, 
where, in particular, the dimension 6 operator 
given in Eq.~\eqref{eq:NBFPortalLagrangian} 
has been analysed.
Indirect searches at neutrino detectors will always play a sub-leading role as long as annihilation to charged leptons is possible. Indeed, present constraints from DM annihilation to charged leptons are strong enough to rule out the entire allowed region
of the parameter space that could lead to the correct DM relic density as long as the coupling to electrons is sizeable. However, if DM dominantly couples to the heavier lepton generations, allowed windows open up for $m_\chi < m_\mu~(m_\tau)$ (see 
the upper-right and bottom-left panels of Fig.~\ref{Fig:BruteForceResults}). 
In this case, the DM relic density would be set by its annihilation to neutrinos, and the most relevant present constraints come from the results of SK and Borexino. The prospects for HK and DUNE would be very promising in these scenarios, allowing to probe most of the parameter space up to and beyond where the relic density is entirely explained by freeze-out based on neutrino interactions.  

Regarding the constraints that could be set by the DM effects in the spectrum or isotropy of high energy cosmic neutrinos as observed by IceCube~\cite{Arguelles:2017atb}, these would lie in the region of the parameter space already excluded by the number of relativistic degrees of freedom in the early Universe~\cite{Boehm:2013jpa,Nollett:2014lwa,Escudero:2018mvt}.

From Fig.~\ref{Fig:BruteForceResults} it is clear that, 
as long as light DM couples to the electron doublet, this option for a neutrino-DM coupling is mostly ruled out by DM-electron interactions. 
However, if the DM coupling to $L_e$ 
is negligible and DM dominantly couples to $L_\mu$ and/or $L_\tau$, the viable part of parameter space with $m_\chi<m_\mu~(m_\tau)$ 
can be probed by the neutrino experiments.

\section{Coupling via the neutrino portal}
\label{sec:NPortal}
Given the results of the previous section, we will now explore whether the neutrino portal option is able to lead to a rich DM-neutrino phenomenology without being in conflict with indirect searches involving charged leptons. The first necessary ingredient is to have sizeable mixing between the SM neutrinos and the new sterile neutrinos that will mediate the DM interaction.
Therefore, the sterile-light neutrino mixing should not scale with the light neutrino masses, unlike in the canonical seesaw mechanism. Therefore, we will instead attribute the smallness of neutrino masses to an approximate lepton number (or $B-L$) symmetry rather than to a hierarchy of scales between the Dirac and Majorana masses. The new singlets will thus form pseudo-Dirac pairs since lepton number violation will necessarily be very small to account for the lightness of SM neutrinos. This is the case for instance in the popular ``inverse''~\cite{Mohapatra:1986bd,Bernabeu:1987gr} and ``linear''~\cite{Malinsky:2005bi} seesaw mechanisms based on such a symmetry.

As a simplifying assumption we will here consider the addition of only one (pseudo-)Dirac sterile neutrino that will serve as portal between the SM neutrinos and DM. Neglecting this small lepton number violation, the couplings between the SM and the new Dirac singlet neutrino are given by
\begin{align}
\mathcal{L} &= \mathcal{L}_\mathrm{SM}  
+ \overline{N}\left(i\slashed{\d} - m_N\right)N 
- \l_\a \overline{L_\a} \tilde{H} N_R\,,
\label{eq:NPortal}
\end{align}
where  
$N$ is the Dirac sterile neutrino and $\tilde{H} = i \sigma_2 H^*$, with $H$ being the Higgs doublet. 

Electroweak symmetry breaking gives rise to the neutrino Dirac mass term 
\be
\left(\overline{\nu_{\a L}},\, \overline{N_L}\right) M_\nu N_R + \mathrm{h.c.}\,,
\ee
where $M_\nu = (\l_\a v ,\, m_N)^T$ is the neutrino mass matrix
and $v = \ev{H^0} = 174$~GeV is the Higgs vacuum expectation value (vev).
Diagonalising $M_\nu M_\nu^\dagger$ with a $4\times 4$ unitary matrix $U$, 
\be
U^\dagger M_\nu M_\nu^\dagger\, U = \diag\left(m^2_1, m^2_2, m^2_3, m^2_4\right),
\ee
we find the mass of the heavy neutrino to be
\be
m_4 = \sqrt{m_N^2 + \sum_{\a} |\l_\a|^2 v^2}\,.
\ee
As expected, the lepton number symmetry forbids light neutrino masses. In order to account for neutrino masses, small breaking of this symmetry via terms such as $\mu\, \overline{N_L} N^c_L$ (inverse seesaw), or $\lambda_\alpha' \overline{L_\a} \tilde{H} N^c_L $ (linear seesaw) can be added. Since these small parameters would have negligible impact in the phenomenology of neutrino-DM interactions, we will not consider them in what follows.

The neutrino mixing matrix $U$, 
which relates LH flavour neutrino fields and 
the neutrino fields with definite masses as 
\be
\begin{pmatrix}
\nu_{\a L} \\
N_L
\end{pmatrix} 
= U \begin{pmatrix}
\nu_{iL} \\
\nu_{4L}
\end{pmatrix},
\quad
\a = e,\mu,\tau\,,
\quad
i = 1,2,3\,,
\ee
 has the form
\be
U = \begin{pmatrix}
U_{\a i} & U_{\a 4} \\
U_{s i} & U_{s4}
\end{pmatrix}.
\ee
%
The upper-left $3\times3$ block $U_{\a i}$ would
correspond to the Pontecorvo--Maki--Nakagawa--Sakata (PMNS) matrix once the small lepton number-breaking terms that induce neutrino masses are taken into account. Note that this matrix, being a $3 \times 3$ sub-block of a larger unitary matrix will, in general, not be unitary.   
The upper-right $3\times1$ block $U_{\a 4}$ describes the mixing between 
the active flavour neutrinos and the LH component 
of the heavy neutrino with mass $m_4$. 
The last row of the matrix $U$ specifies the admixture of each $\nu_{jL}$, $j=1,2,3,4$, in 
the LH sterile neutrino $N_L$. As we will see in what follows, the DM-related phenomenology 
is driven by the mixing of active-heavy mixing matrix elements $U_{\a4}$.
We will use the unitarity deviations of the PMNS matrix to constrain these mixings~\cite{Fernandez-Martinez:2016lgt}.
The mixing elements of interest are given by 
\be
U_{\a4} = \frac{\th_\a}{\sqrt{1 + \sum_\a |\th_\a|^2}} \,,
\qquad
U_{s4} = \frac{1}{\sqrt{1 + \sum_\a |\th_\a|^2}}\,,
\qquad
\sum_{i=1}^3 |U_{si}|^2 = \sum_{\alpha=e}^\tau |U_{\a 4}|^2\,,
\ee
%
with $\theta_\alpha = \lambda_\alpha v/m_N$. Note that, even though the SM neutrino masses have been neglected, the mixing with the extra singlet neutrino that will act as portal can still be sizeable. 
For definiteness we will fix the mixing to the different flavours to their $1\sigma$ limit from Ref.~\cite{Fernandez-Martinez:2016lgt}, namely:
\be
|\theta_e| = 0.031\,,
\quad
|\theta_{\mu}| = 0.011\,,
\quad
|\theta_{\tau}| = 0.044\,.
\ee
%

In the following sections, we will explore two possible ways in which these Dirac neutrinos could couple to the dark sector and become portals between it and the SM neutrinos.

\section{Neutrino portal with a scalar mediator}
\label{sec:Scalar}
In this first example, we will assume that DM is composed of a new fermion, singlet under the SM gauge group, and that a new scalar mediates the Dirac neutrino-DM interactions.

\subsection{Model}
 The Lagrangian of the model we will consider is given by 
\begin{align}
\mathcal{L} &= \mathcal{L}_\mathrm{SM} 
+ \overline{\chi}\left(i\slashed{\d} - m_\chi \right)\chi 
+ \overline{N}\left(i\slashed{\d} - m_N\right)N 
+ \d_\mu S^* \d^\mu S \nn\\
&\phantom{{}={}} - \left[\l_\a \overline{L_\a} \tilde{H} N_R 
+ \overline{\chi} \left(y_L N_L + y_R N_R\right) S + \mathrm{h.c.}\right] \nn\\
&\phantom{{}={}}  - \mu_S^2 |S|^2 - \lambda_S |S|^4 - \lambda_{SH} |S|^2 H^\dagger H \,,
\label{eq:NSPortalLagrangian}
\end{align}
where $\chi$ is a Dirac fermion DM candidate and $S$ is a complex scalar. The fields $\chi$ and $S$ form the dark sector of the model (they are SM singlets), while $N$ serves as a mediator between the dark sector and SM. 
The Lagrangian in Eq.~\eqref{eq:NSPortalLagrangian} respects a global $U(1)_L$ lepton number symmetry under which $L_\a$, $N$, and $S^*$ have the same charge and which protects the SM neutrino masses. 
Moreover, the Lagrangian respects a global $U(1)_D$ dark symmetry, under which $\chi$ and $S$ have equal charges. This preserved symmetry ensures the stability of $\chi$, if $m_\chi < m_S$, where $m_S^2 = \mu_S^2 + \l_{SH}v^2$ is the mass squared of the scalar $S$. 
For $m_\chi > m_S$, the roles of $\chi$ and $S$ would change, 
and $S$ would be a DM candidate. 
While this possibility is perfectly viable, 
it is more difficult to probe at neutrino detectors, 
as the DM annihilation cross section to neutrinos 
is velocity-suppressed. 
In what follows we assume that $m_\chi < m_S$ and focus on fermionic DM.

This model was previously considered in Refs.~\cite{Bertoni:2014mva,Batell:2017cmf}.
However, we will go beyond these works by performing a comprehensive analysis of the sensitivity of neutrino experiments to the parameter space of this model. 

We will limit ourselves to the case in which DM is lighter than the heavy neutrino,%
\footnote{Otherwise the $\chi \overline{\chi} \to \nu_i \overline{\nu_4}$ or $\chi \overline{\chi} \to \nu_4 \overline{\nu_4}$ channels would dominate the annihilation cross section and only sub-dominant DM interactions with the 3 light SM neutrinos $\nu_i$ would be allowed.} 
i.e., $m_\chi < m_4$. This is the so-called direct annihilation regime \cite{Pospelov:2007mp}, 
since DM annihilates through the mediator directly to SM particles. 
As intended, the only channel for DM annihilation at tree level is 
the one into light neutrinos. 
This process occurs via a diagram involving a $t$-channel exchange of 
the scalar mediator $S$. 
In the opposite regime, 
which is usually referred to as secluded~\cite{Pospelov:2007mp}, 
DM annihilates to heavy neutrinos, 
which subsequently decay. 
The phenomenology of this regime has been studied 
in Refs.~\cite{Escudero:2016tzx,Escudero:2016ksa,Folgado:2018qlv,Bandyopadhyay:2018qcv}.

Neglecting velocity-suppressed terms, we find 
the following thermally averaged 
cross section for DM annihilation to neutrinos:
\be
\ev{\s\vr} \approx
\frac{y_L^4 }{32\pi}\left(\sum_{i = 1}^3 |U_{s i}|^2\right)^2
\frac{m_\chi^2}{\left(m_\chi^2+m_S^2\right)^2} \approx
\frac{y_L^4 }{32\pi}\left(\sum_{\a=e,\mu,\tau} \left|\th_\a\right|^2\right)^2
\frac{m_\chi^2}{\left(m_\chi^2+m_S^2\right)^2}\,.
\label{eq:XSscalar}
\ee
The product $y_L \sqrt{\sum_\a|\th_\a|^2}$ controls $\ev{\s\vr}$
and, in order to allow for sufficient annihilation to reproduce the observed relic density, it cannot be too small. 
The value of the coupling $y_L$ is limited by the requirement of perturbativity. We will restrict ourselves to $y_L < 4\pi$. 
Since the coupling $y_R$ does not 
enter Eq.~\eqref{eq:XSscalar}, and thus, 
does not affect the tree-level DM-neutrino interactions, 
in what follows we set it to zero for simplicity.
Regarding the mixing parameters $\th_\a$, 
the bounds on them depend on the mass of the heavy neutrino.
For definiteness we will assume that the heavy neutrino has a mass above 
the electroweak scale. At this scale the bounds on heavy neutrino mixing derived in the global analysis of flavour and electroweak precision data 
performed in Ref.~\cite{Fernandez-Martinez:2016lgt} apply. If smaller masses were instead considered, more stringent constraints from collider and beam-dump searches and, eventually, production in meson and beta decays 
could potentially apply~\cite{Atre:2009rg} 
(see discussion in Section~\ref{sec:VectorResults}).
In any case, all the observables relevant to DM phenomenology
have a sub-leading dependence on $m_4$.
We also consider the case where the coupling $\l_{SH} = 0$,
ensuring the neutrino portal regime. 
In Refs.~\cite{Bertoni:2014mva,Batell:2017cmf}, the radiative generation of the $|S|^2 H^\dagger H$ operator 
was considered and its effects on $m_S$ as well as on the invisible width of the Higgs boson were found to be negligible.
\begin{figure}
\centering
\includegraphics[width=10cm]{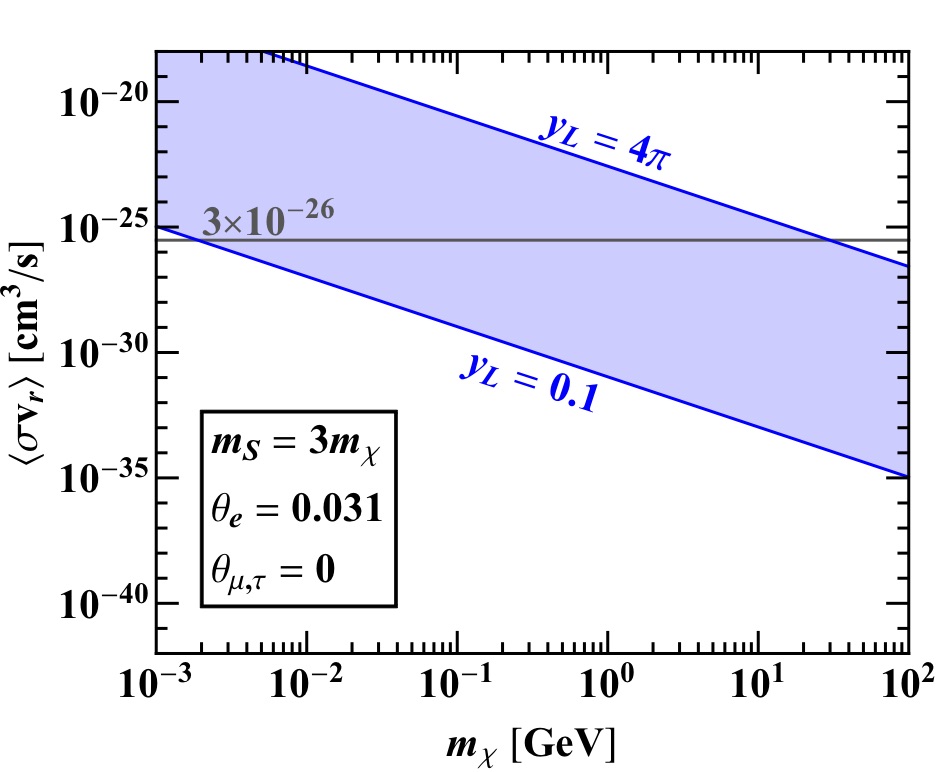}
\caption{Thermally averaged annihilation cross section multiplied by the relative velocity 
for $\chi\overline{\chi} \to \nu \overline{\nu}$. 
We have fixed $m_S = 3m_\chi$, $\th_e = 0.031$, 
$\th_\mu = \th_\tau = 0$,
and varied $y_L$ between $0.1$ and $4\pi$.}
\label{fig:mDMsigvNuNuEmix}
\end{figure}
%

In Fig.~\ref{fig:mDMsigvNuNuEmix}, we
show the region of the parameter space for which the correct thermal relic abundance is obtained. This region spans DM masses up to $100$~GeV for $|\th_e| = 0.031$, $\th_\mu = \th_\tau = 0$, 
and $y_L$ between $0.1$ and $4\pi$ while
keeping $m_S = 3m_\chi$ as a benchmark.

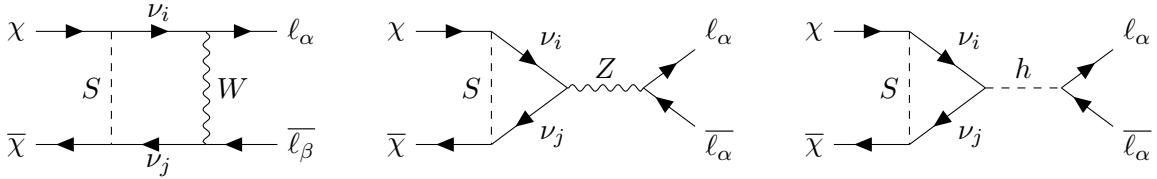
\begin{figure}
\centering
\begin{tikzpicture}
\begin{feynman}
  \vertex (1) at (-3.75,.75) {$\chi$};
  \vertex (a1) at (-2.5, 0.75);
  \vertex (b1) at (-1.25, 0.75);
  \vertex (3) at (0, 0.75) {$\ell_\alpha$};
  \vertex (2) at (-3.75, -0.75) {$\overline{\chi}$};
  \vertex (a2) at (-2.5, -0.75);
  \vertex (b2) at (-1.25, -0.75);
  \vertex (4) at (0, -0.75) {$\overline{\ell_\beta}$};
  \vertex (1a) at (1.25, 0.75) {$\chi$};
  \vertex (aa1) at (2.5, 0.75);
  \vertex (ba1) at (3.5, 0);
  \vertex (3a) at (5.5, 0.75) {$\ell_\alpha$};
  \vertex (2a) at (1.25, -0.75) {$\overline{\chi}$};
  \vertex (aa2) at (2.5, -0.75);
  \vertex (ba2) at (4.5, 0);
  \vertex (4a) at (5.5, -0.75) {$\overline{\ell_\alpha}$};
  \vertex (1ab) at (6.75, 0.75) {$\chi$};
  \vertex (aab1) at (8, 0.75);
  \vertex (bab1) at (9, 0);
  \vertex (3ab) at (11, 0.75) {$\ell_\alpha$};
  \vertex (2ab) at (6.75, -0.75) {$\overline{\chi}$};
  \vertex (aab2) at (8, -0.75);
  \vertex (bab2) at (10, 0);
  \vertex (4ab) at (11, -0.75) {$\overline{\ell_\alpha}$};
  \diagram*
  {
  (1) -- [fermion] (a1),
  (a1) -- [fermion, edge label=$\nu_i$] (b1),
  (b1) -- [fermion] (3),
  (a1) -- [scalar, edge label'=$S$] (a2),
  (2) -- [anti fermion] (a2),
  (b1) -- [boson, edge label=$W$] (b2),
  (a2) -- [anti fermion, edge label'=$\nu_j$] (b2),
  (b2) -- [anti fermion] (4)
  (1a) -- [fermion] (aa1),
  (aa1) -- [fermion, edge label=$\nu_i$] (ba1),
  (2a) -- [anti fermion] (aa2),
  (aa2) -- [anti fermion, edge label'=$\nu_j$] (ba1),
  (ba1) -- [boson, edge label=$Z$] (ba2),
  (ba2) -- [fermion] (3a),
  (ba2) -- [anti fermion] (4a),
  (aa1) -- [scalar, edge label'=$S$] (aa2)
  (1ab) -- [fermion] (aab1),
  (aab1) -- [fermion, edge label=$\nu_i$] (bab1),
  (2ab) -- [anti fermion] (aab2),
  (aab2) -- [anti fermion, edge label'=$\nu_j$] (bab1),
  (bab1) -- [scalar, edge label=$h$] (bab2),
  (bab2) -- [fermion] (3ab),
  (bab2) -- [anti fermion] (4ab),
  (aab1) -- [scalar, edge label'=$S$] (aab2)
  };
  \end{feynman}
  \end{tikzpicture}
  \caption{One-loop diagrams (in unitary gauge) 
contributing to 
annihilation of DM into charged lepton-antilepton pairs $\ell_\alpha \overline{\ell_\beta}$, $\alpha,\beta = e,\mu,\tau$. The indices $i$ and $j$ run from 1 to 4.}
  \label{fig:chichitoll}
  \end{figure}

Annihilation of DM into charged lepton-antilepton pairs $\ell^+\ell^-$ 
($\ell = e,\mu,\tau$) proceeds via the one-loop diagrams%
\footnote{The Feynman diagrams in this article are produced 
with the \texttt{Ti\emph{k}Z-Feynman} package \cite{Ellis:2016jkw}.}
shown in Fig.~\ref{fig:chichitoll} (in unitary gauge).

The dominant contribution comes from the first and second diagrams, 
while the contribution from the last diagram is suppressed by the small 
Yukawa couplings of the charged leptons. 
The first diagram leads to the following effective operator:
\be
\mathcal{L} \supset -a_{SW}\frac{g^2}{m_W^2}
\overline{\chi} \g^\mu P_R \chi\, 
\overline{\ell_\alpha} \g_\mu P_L \ell_\beta\,,
\ee
where $g$ is the weak coupling constant. 
Neglecting external momenta, the effective coupling $a_{SW}$ is given by
\be
a_{SW} = |U_{s4}|^2 U_{\alpha 4} U^*_{\beta 4} 
\frac{y_L^2}{(4\pi)^2} G\left(\frac{m_S^2}{m_4^2}\right),
\ee
where the loop function $G(x)$ reads
\be
G(x) = \frac{x - 1 - \log{ x}}{4 \left(1-x\right)^2}\,.
\ee
%
The second diagram in Fig.~\ref{fig:chichitoll} leads to the following effective interaction of DM with the $Z$ boson:
\be
\mathcal{L} \supset - a_Z \frac{g}{\cos\th_W} 
\overline{\chi} \g^\mu P_R \chi Z_\mu \,,
\ee
%
where $\th_W$ is the Weinberg angle and 
$a_Z$ is the effective coupling, which
in the limit of zero external momenta is given by
\be
a_Z = |U_{s4}|^2 \left(1 - |U_{s4}|^2\right) 
\frac{y_L^2}{(4\pi)^2} G\left(\frac{m_S^2}{m_4^2}\right)\,.
\label{eq:aZ}
\ee
%

%
%

These contributions have been also computed
using a combination of packages:
\texttt{FeynRules}~\cite{Christensen:2008py,Alloul:2013bka} to produce a model file, 
\texttt{FeynArts}~\cite{Hahn:2000kx} for generating the diagrams and 
\texttt{FormCalc}~\cite{Hahn:1998yk} for computing their numerical contributions.
For numerical evaluation of the Passarino-Veltman functions 
we have used \texttt{LoopTools}~\cite{Hahn:1998yk}. 
We have also considered the limit of zero external momenta, 
which effectively corresponds to the limit of small DM and charged lepton masses, 
and confronted the analytical results obtained in this approximation using 
the package \texttt{ANT} \cite{Angel:2013hla} with the \texttt{LoopTools} results. For DM masses between $1$~MeV and $100$~GeV that we are interested in, 
the approximation works very well. The availability of analytical expressions 
allows for an easier exploration of the parameter space.

\begin{figure}
\centering
\includegraphics[width=7.6cm]{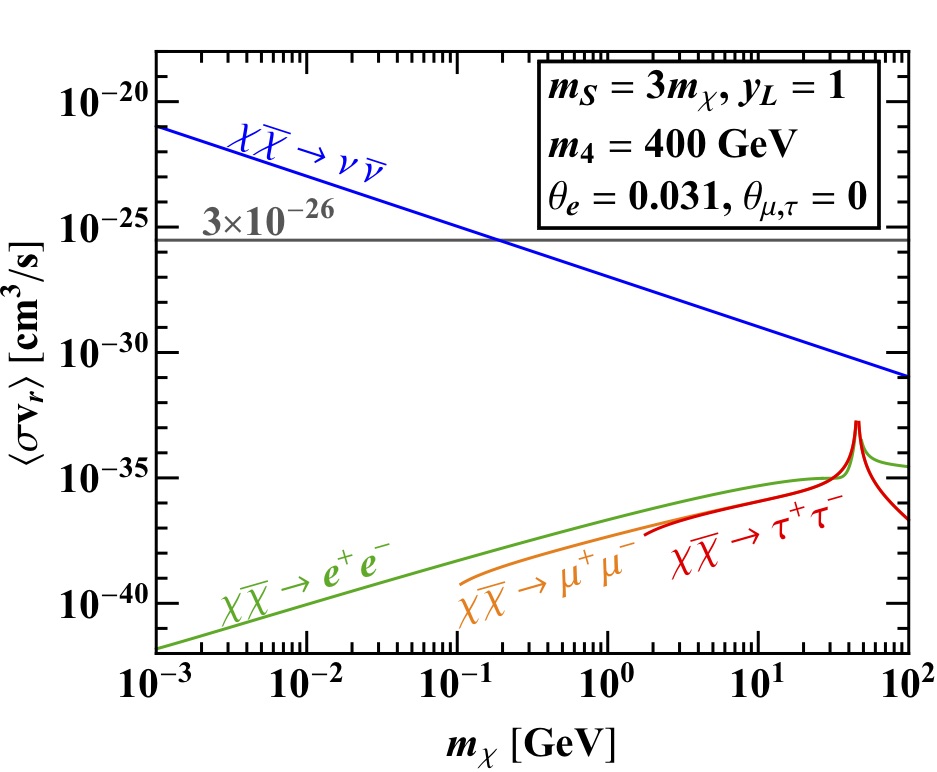}
\hspace{0.1cm}
\includegraphics[width=7.6cm]{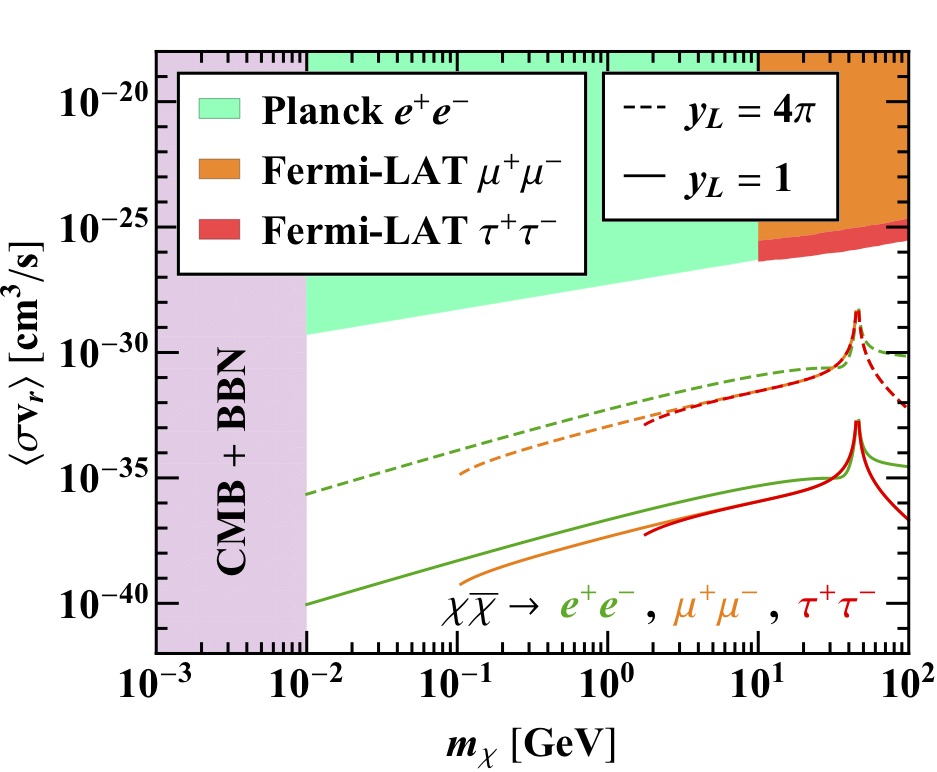}
\caption{Thermally averaged annihilation cross section multiplied by the relative velocity 
for DM annihilation into $e^+e^-$, $\mu^+\mu^-$, and $\tau^+\tau^-$. 
We have fixed $m_S = 3m_\chi$, $m_4 = 400$~GeV, 
$y_L = 1$, $\th_e = 0.031$, and $\th_{\mu,\tau} = 0$. 
The \textit{left} panel provides comparison with $\ev{\s\vr}$ for DM annihilation into neutrinos assuming the same set of model parameters. 
The \textit{right} panel displays the indirect detection constraints 
coming from Planck and Fermi-LAT. 
The lower bound $m_\chi \gtrsim 10$~MeV 
is set by observations of the CMB and BBN. See text for further details.}
\label{fig:mDMsigvEmiXyL1}
\end{figure}

In Fig.~\ref{fig:mDMsigvEmiXyL1}, we present the cross sections for annihilation of DM into $e^+e^-$, $\mu^+\mu^-$, and $\tau^+\tau^-$ 
for benchmark values of the model parameters. We fix $m_S = 3m_\chi$, $m_4 = 400$~GeV, 
$y_L = 1$, $\th_e = 0.031$, and $\th_{\mu,\tau} = 0$. 
As can be seen from the left panel, 
the annihilation cross sections to charged leptons
are several orders of magnitude smaller than the cross section 
for DM annihilation into neutrinos. 
The difference in the cross sections becomes smaller when the DM mass approaches $m_Z/2$, 
and the cross sections for 
$\chi \overline{\chi} \to \ell^+ \ell^-$
exhibit a resonant behaviour 
due to the second diagram in Fig.~\ref{fig:chichitoll}. 
In the right panel, we show the indirect detection constraints 
from Planck \cite{Aghanim:2018eyx, Slatyer:2015jla} and Fermi-LAT~\cite{Ahnen:2016qkx}. 
Note that those constraints assume a $100\%$ annihilation rate 
into a single SM channel. 
Even for $y_L = 4\pi$ the resulting annihilation cross sections into charged leptons 
are well below the experimental constraints. 
Thus, the considered realisation of the neutrino portal 
does provide an example of a gauge-invariant model
in which the neutrino-DM interactions dominate 
DM phenomenology.
%

At one-loop level DM also interacts with quarks via 
diagrams involving $Z$ and $h$, 
which are analogous to those in Fig.~\ref{fig:chichitoll}. 
The corresponding effective DM-nucleon spin-independent 
scattering cross section reads~\cite{Batell:2017cmf} 
\be
\s_n = \frac{\mu_n^2}{\pi} \frac{\left(Z f_p + \left(A-Z\right) f_n\right)^2}{A^2}\,,
\ee
%
where $\mu_n$ is the reduced mass of the nucleon, 
$A$ is the total number of nucleons in a nuclei, 
$Z$ is the number of protons,
\be
f_p = \left(4 \sin^2\th_W - 1\right) \frac{G_F a_Z}{\sqrt{2}}\,, 
\qquad 
f_n = \frac{G_F a_Z}{\sqrt{2}}\,,
\ee
%
with $a_Z$ given in Eq.~\eqref{eq:aZ}, 
and $G_F$ being the Fermi constant.
The radiative coupling of DM to the Higgs, $\overline{\chi} \chi h$, would also give a contribution to direct detection searches. This contribution is however suppressed  by the small quark Yukawa couplings.
Direct detection of a SM singlet 
fermion DM candidate at one loop has been recently studied in detail in \cite{Herrero-Garcia:2018koq}. Moreover, an interesting example, which also provides 
radiative generation of neutrino masses, has been presented in \cite{Hagedorn:2018spx}.

The most stringent constraint on DM-nucleon 
spin-independent cross section 
for $m_\chi \gtrsim 10$~GeV comes from XENON1T \cite{Aprile:2018dbl}. 
As we will see in the next subsection, this constraint is strong enough 
to probe the loop-suppressed scattering process
if the value of the coupling $y_L$ is sufficiently large.
We have also considered DM scattering off electrons and found 
that the corresponding cross section is much smaller than 
the projected sensitivities of silicon, germanium, and xenon experiments 
derived in Ref.~\cite{Essig:2018tss}. 
Thus, DM-electron scattering cannot provide an additional probe 
of the considered neutrino portal model.

\subsection{Results}
\begin{figure}
\vspace{-3cm}
\centering
\includegraphics[width=7.5cm]{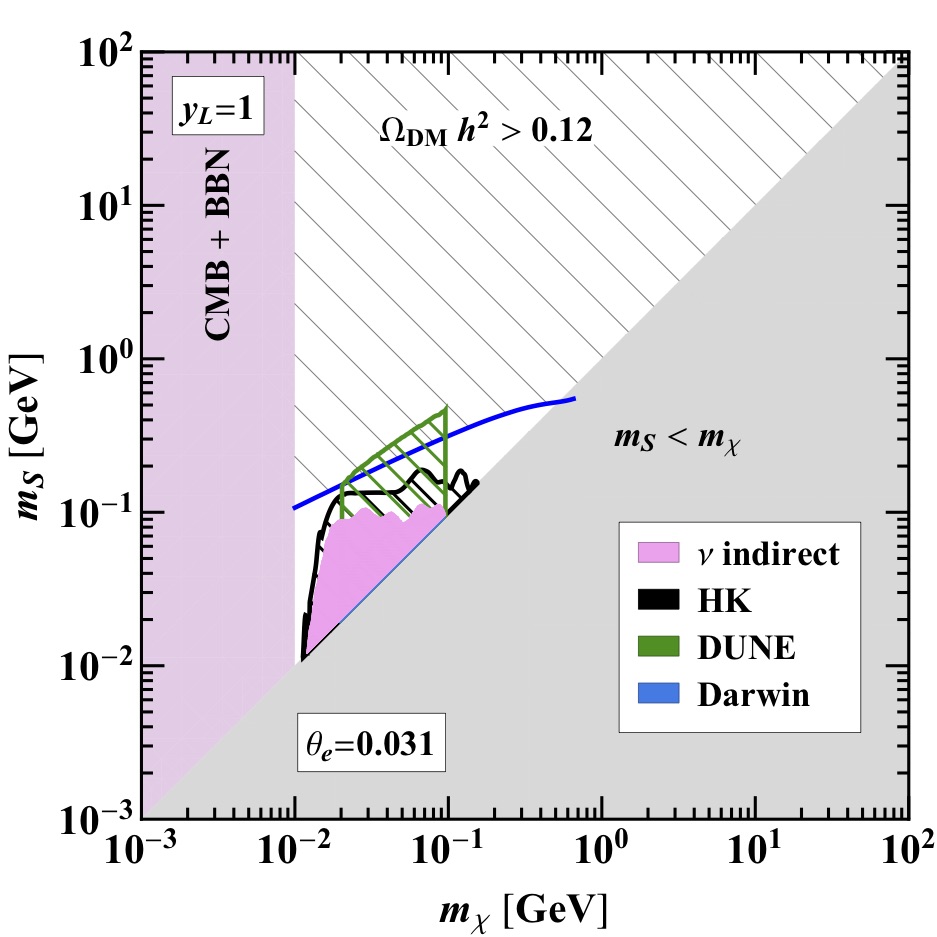}
\hspace{0.1cm}
\includegraphics[width=7.5cm]{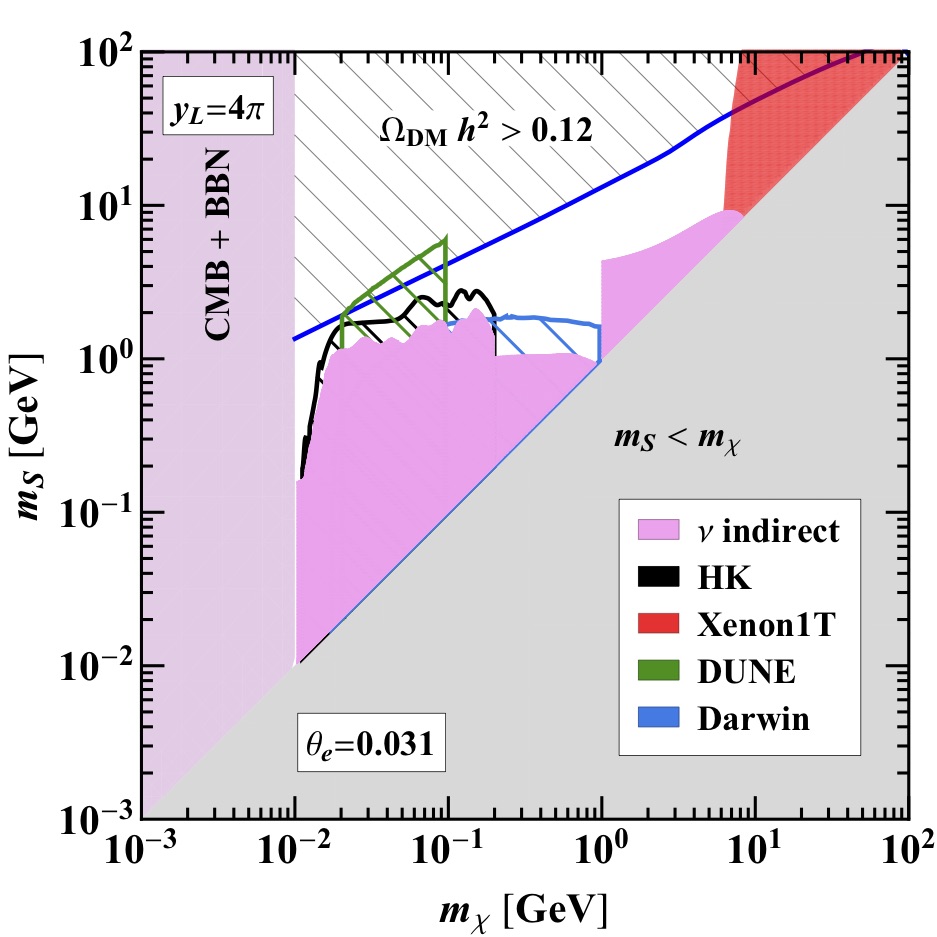}
\includegraphics[width=7.5cm]{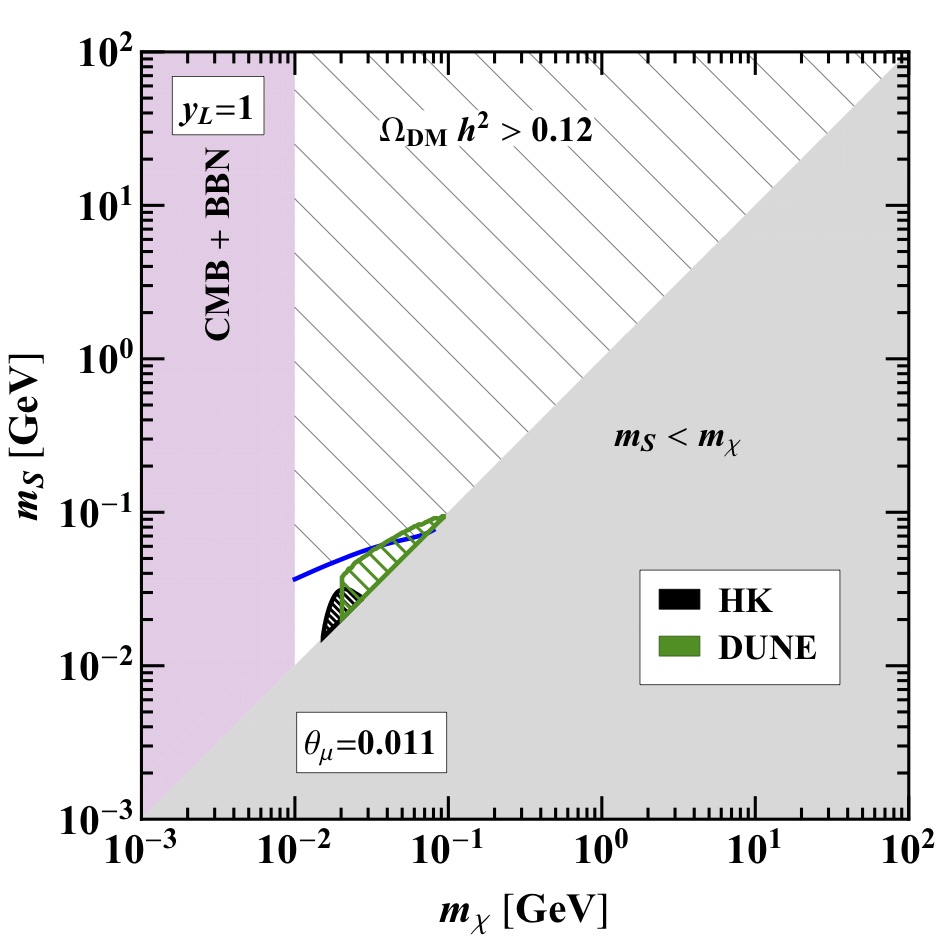}
\hspace{0.1cm}
\includegraphics[width=7.5cm]{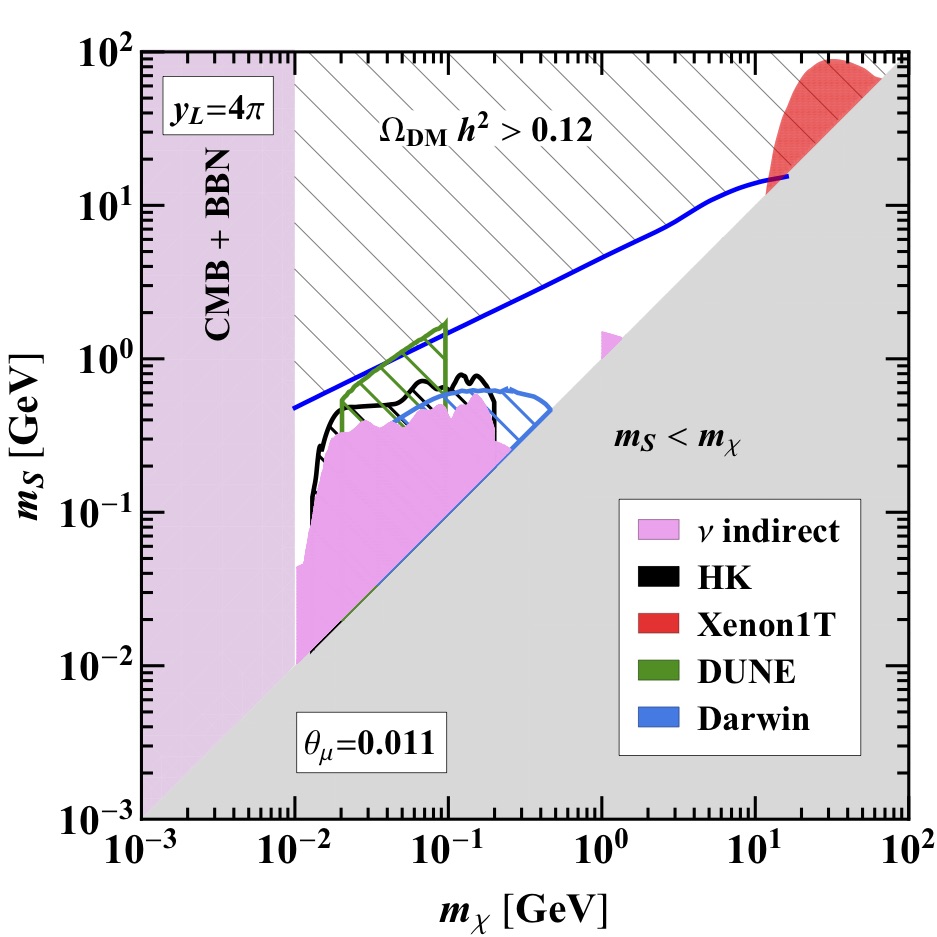}
\includegraphics[width=7.5cm]{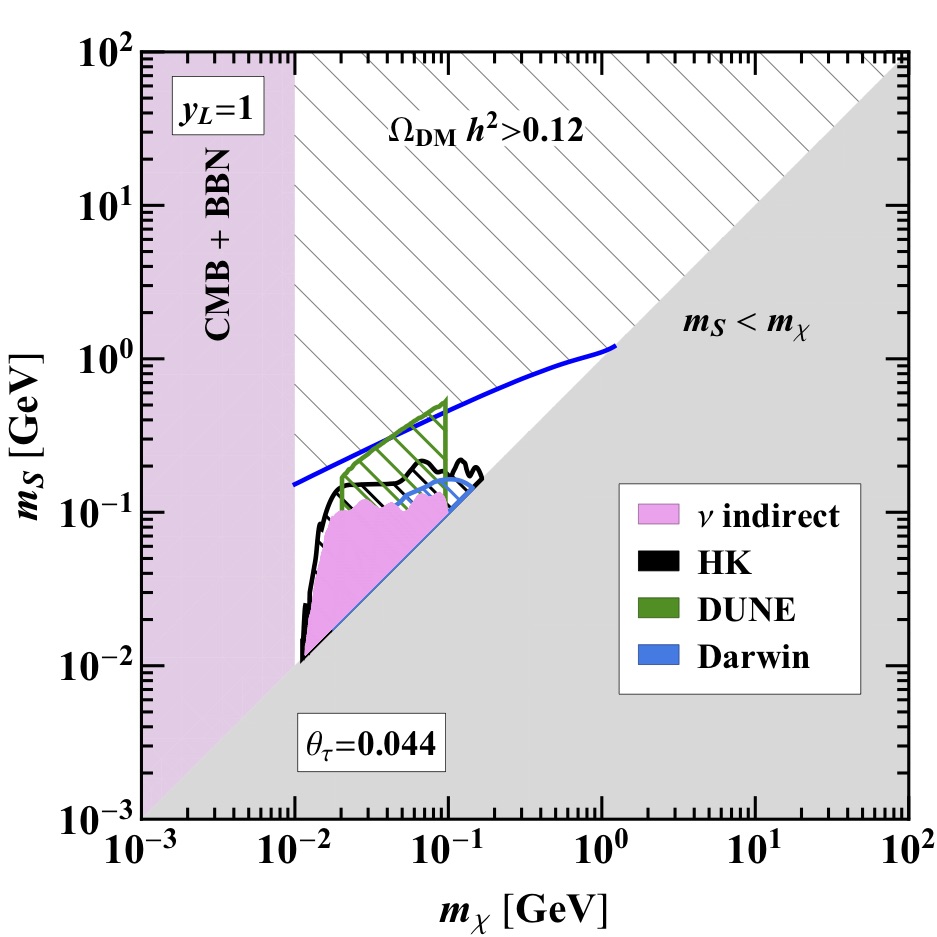}
\hspace{0.1cm}
\includegraphics[width=7.5cm]{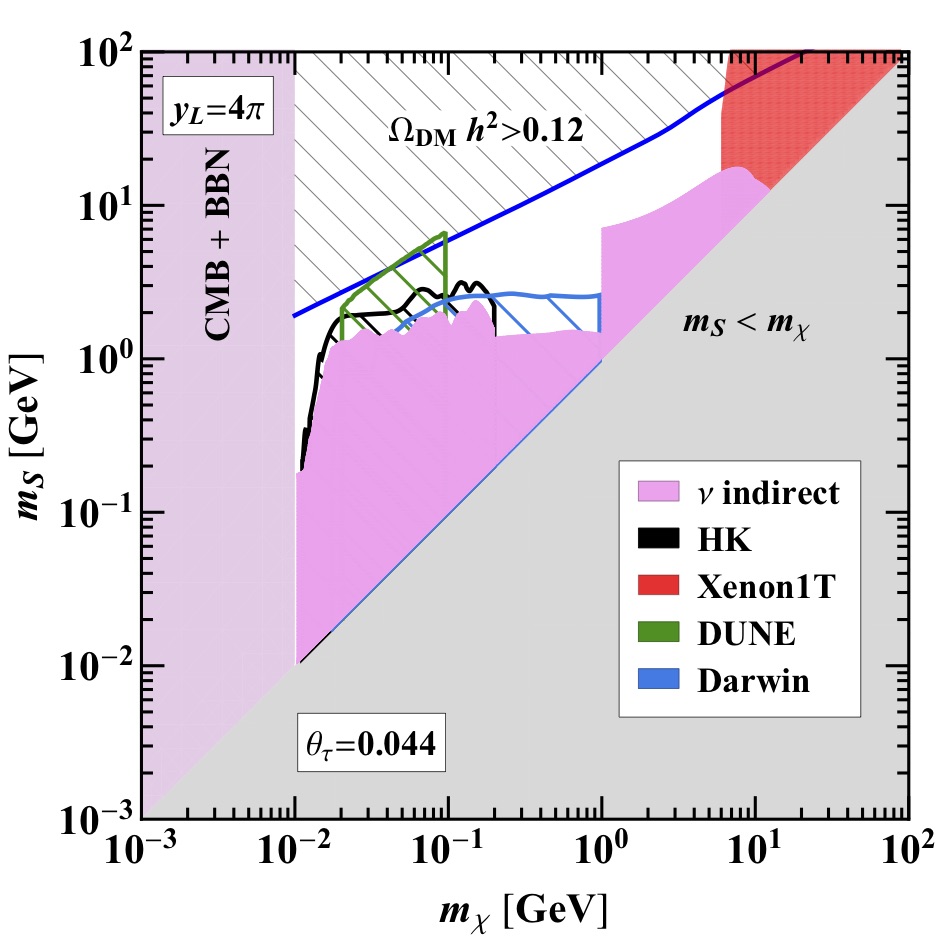}
\caption{Constraints on the DM mass $m_\chi$ and the dark scalar mass $m_S$. 
We have fixed $\th_e = 0.031$, $\th_{\mu,\tau} = 0$; $\th_\mu = 0.011$, $\th_{e,\tau} = 0$; and $\th_\tau = 0.044$, $\th_{e,\mu} = 0$ (from top to bottom), 
considering $y_L = 1$ and $4\pi$. 
Along the blue line  
the DM relic density matches the observed value.
The coloured shaded regions are excluded by different experiments, while the hatched areas correspond to prospective sensitivities of future experiments.
The lower bound $m_\chi \gtrsim 10$~MeV 
is set by observations of the CMB and BBN. See text for further details.}
\label{fig:mDMmS}
\end{figure}

In this subsection, we explore the parameter space to find regions that satisfy all direct and indirect detection constraints and in which the DM phenomenology could be dominated by its interactions with SM neutrinos.
We show our results in
the $m_\chi$-$m_S$ plane  
to determine
the masses of the DM and the dark scalar that are presently allowed and could lead to the correct relic abundance 
(see Fig.~\ref{fig:mDMmS}).  
%

 In Fig.~\ref{fig:mDMmS}, the triangular region $m_S < m_\chi$ 
is forbidden by DM stability.
Along the blue line(s) computed with \texttt{micrOMEGAs},%
\footnote{We have implemented the effective DM couplings to the $Z$ boson 
and to the charged leptons via exchange of the $W$ boson (see Fig.~\ref{fig:chichitoll}) 
to the \texttt{FeynRules} model file.} 
the DM relic density matches the observed value  
$\O_\mathrm{DM} h^2 = 0.1193 \pm 0.0009$~\cite{Aghanim:2018eyx}. 
Above this line (the upper hatched region), 
the DM relic density is bigger than the measured value, 
i.e., DM overcloses the Universe. 
Below this line, the relic abundance would be smaller than the observed value.
However, if there is an additional production mechanism, 
the relic abundance could also be compatible with this region.

As can be seen in the figure, indirect searches for annihilation to neutrinos, together with direct detection bounds by XENON1T for large DM masses, are the only probes that are presently constraining the allowed parameter space. The prospects to explore the remaining allowed regions through annihilation to neutrinos are very promising. In particular DUNE would be able to detect the neutrino signal in the range $25-100$~MeV if the DM abundance is entirely due to this process.

\begin{figure}
\centering
\includegraphics[width=10cm]{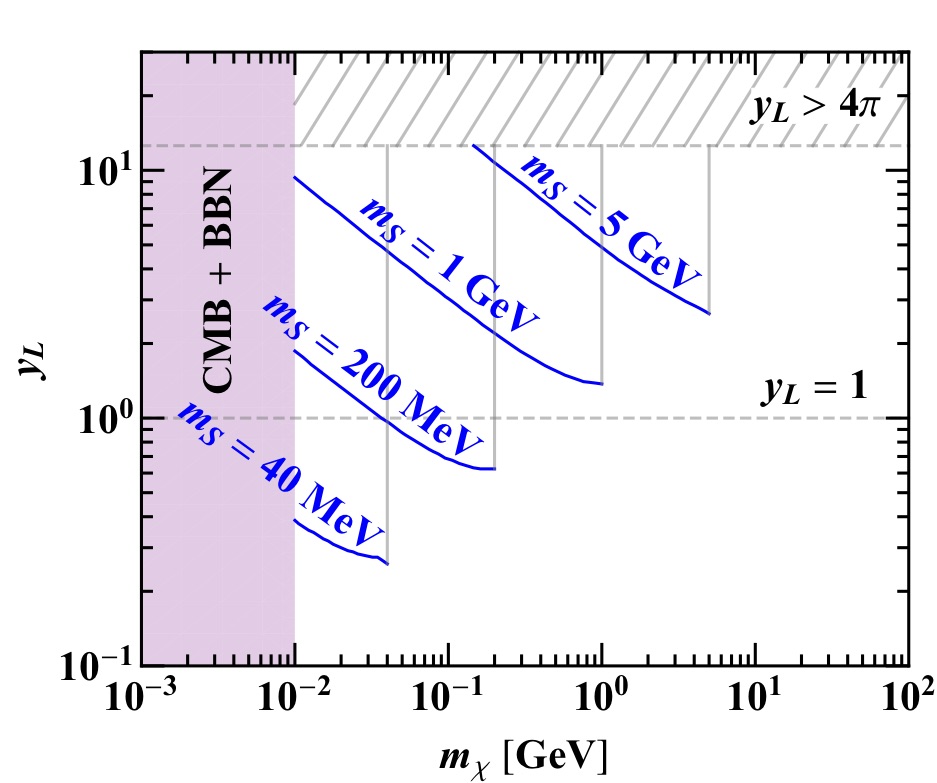}
\caption{Values of the DM mass $m_\chi$ and the coupling $y_L$ required to reproduce the
observed relic abundance.
We have fixed $m_S = 0.04$, $0.2$, $1$, and $5$~GeV, 
and have considered the representative case of $\th_e = 0.031$, while keeping $\th_{\mu,\tau} = 0$.
Along (above) the blue lines 
the DM relic density matches 
(is less than) the observed value. 
The lower bound $m_\chi \gtrsim 10$~MeV 
is set by observations of the CMB and BBN.}
\label{fig:mDMyLeMiX}
\end{figure}

In Fig.~\ref{fig:mDMyLeMiX}, we fix $m_S$ to several representative values, 
namely $m_S = 0.04$, $0.2$, $1$, and $5$~GeV, and 
show the lines corresponding to the correct relic abundance 
in the $m_\chi$-$y_L$ plane. 
These results have been obtained with \texttt{micrOMEGAs}.
Small values of $y_L$ are ruled out since they do not lead to 
efficient DM annihilation. As can be seen, a lighter dark scalar allows for 
smaller values of $y_L$. For $m_S \gtrsim 500$~MeV, the values of $y_L \gtrsim 1$ are required to yield the observed relic density.
%

Overall, the cosmologically allowed parameter space of the model is 
already constrained by the current neutrino detectors as well as XENON1T.%
\footnote{For $m_\chi>5$~GeV, DARWIN 
will have a better sensitivity to spin-independent 
DM-nucleon cross section than that of XENON1T~\cite{Aalbers:2016jon}.
However, for $y_L = 4\pi$, these masses are already ruled by XENON1T, 
while for $y_L=1$, they are not allowed by 
the relic abundance constraint.}
Moreover, the next generation of neutrino experiments, in particular DUNE, 
will be able to probe thermal MeV fermion DM in the considered scenario.

\section{Neutrino portal with a vector mediator}
\label{sec:Vector}
In this second example, we will again assume that DM is composed of a new Dirac fermion, this time coupled to a new 
massive vector boson. The Dirac singlet neutrino will also interact with this boson so as to provide the neutrino-DM interaction.

\subsection{Model}
The Lagrangian of the model is given by
\begin{align}
\nonumber
\mathcal{L} &= \mathcal{L}_\mathrm{SM} 
+ \overline{\chi}\left(i\slashed{\d} - m_\chi \right)\chi 
+ \overline{N}\left(i\slashed{\d} - m_N\right)N \\
\nonumber
&\phantom{{}={}} + g' \overline{\chi_R} \gamma^{\mu}\chi_R Z'_{\mu} + g' \overline{N_L} \gamma^{\mu} N_L Z'_{\mu}
-\left[\lambda_{\alpha}\overline{L_{\alpha}}\tilde{H}N_R + \mathrm{h.c.} \right] \\
&\phantom{{}={}} -\frac{1}{4} Z'_{\mu\nu}Z'^{\mu\nu} +\frac{1}{2}m_{Z'}^2 Z'_{\mu}Z'^{\mu}\,,
\label{eq:NVPortalLagrangian}
\end{align}
where $\chi$ is a Dirac fermion DM candidate, $Z'$ is a new vector boson mediating the interaction between neutrinos and DM, and $N$ is the Dirac sterile neutrino connecting the dark and visible sectors through its mixing with the active neutrinos. 
This Lagrangian could for instance describe a new $U(1)'$ gauge symmetry spontaneously broken by the vev of a scalar SM singlet charged under it, that would induce masses for the $Z'$ as well as for the heavy neutrino $N$ and the DM. The particular mechanism is not relevant for the rest of the discussion and will not be elaborated further. We will also assume there is an additional conserved charge (e.g., a $\mathbb{Z}_2$ symmetry) not shared between the neutrino and the DM that prevents their mixing.
Note that in order to keep the Lagrangian in Eq.~\eqref{eq:NVPortalLagrangian} anomaly free without introducing new fields, the simplest option is to couple the LH part of the Dirac sterile neutrino and the RH part of the DM to the new gauge boson with the same coupling $g'$.

As in the previous scenario, we will assume that the DM mass $m_{\chi}<m_4$ so that the dominant DM annihilation channel 
is to the three light SM neutrinos. This is a tree-level process and its cross section is given by
\be
\left\langle \sigma v_r\right\rangle \approx \frac{g'^4}{8\pi}\left(\sum_{i = 1}^3 |U_{s i}|^2\right)^2 \frac{m_{\chi}^2}{(4m_{\chi}^2-m_{Z'}^2)^2} \approx \frac{g'^4}{8\pi}\left(\sum_{\a=e,\mu,\tau} \left|\th_\a\right|^2\right)^2 \frac{m_{\chi}^2}{(4m_{\chi}^2-m_{Z'}^2)^2}\,.
\label{eq:AnnCSVector}
\ee
%

\begin{figure}
\centering
\includegraphics[width=10cm]{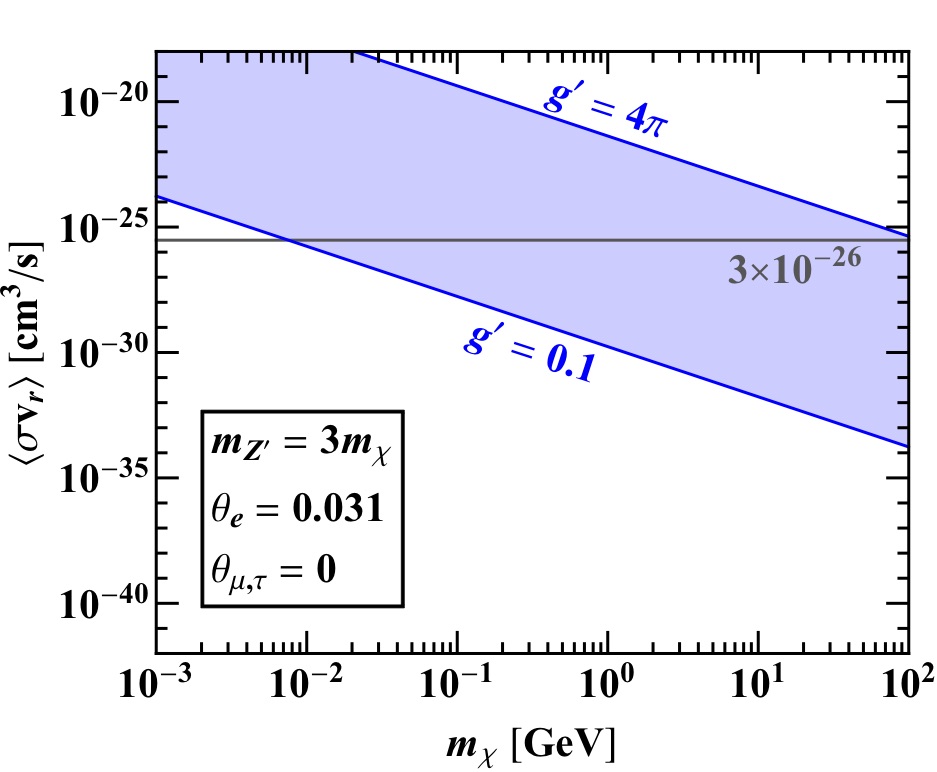}
\caption{Thermally averaged annihilation cross section multiplied by the relative velocity 
for $\chi\overline{\chi} \to \nu \overline{\nu}$. 
We have fixed $m_{Z'} = 3m_\chi$, $\th_e = 0.031$, 
$\th_\mu =\th_\tau=0$, 
and varied $g'$ between $0.1$ and $4\pi$.}
\label{fig:mDMsigvNuNuEmixV}
\end{figure}
%

Note however that, for $m_{Z'} \lesssim m_{\chi}$, the tree-level DM annihilation to a pair of $Z'$ bosons is allowed. When this channel is open, it will dominate
over the direct annihilation into neutrinos, since the latter is suppressed by neutrino mixing. 
This is the so-called secluded annihilation regime \cite{Pospelov:2007mp}, 
which we do not consider in the present study.

In this scenario, as can be seen from Fig.~\ref{fig:mDMsigvNuNuEmixV}, 
 the correct relic abundance can be obtained purely from annihilation 
 to the SM neutrinos for values of the new gauge coupling $g'$ between $0.1$ and $4\pi$, and DM masses in the $0.01-100$~GeV range. 
 In this figure, we have fixed
 $m_{Z'} = 3 m_\chi$, $|\th_e| = 0.031$, and $\th_\mu = \th_\tau =0 $ 
 as benchmark values.

A direct coupling between the $Z'$ boson and the charged leptons will also be induced through the loop diagrams in Fig.~\ref{Fig:ZloopW}. 
Neglecting external momenta for the charged leptons, the effective vertex from the first loop diagram is given by
\be
\mathcal{L}\supset - a_W g' 
\overline{\ell_\alpha} \gamma^{\mu} P_L \ell_{\beta} Z'_{\mu}\,,
\label{eq:VertexWZP}
\ee
where
\be
a_W = |U_{s4}|^2 U_{\alpha 4}U_{\beta 4}^* 
\frac{g^2}{(4\pi)^2}\frac{m_4^2}{2m_W^2}\,.
\label{eq:alphaWZP}
\ee

\subsection{Mixing with the $Z$ boson}
\label{sec:Z-Z'mixing}
\begin{figure}
 \centering
 \begin{tikzpicture}
   \begin{feynman}
   \vertex (1) at (-0.5,0) {$Z'$};
   \vertex (a) at (1,0);
   \vertex (b) at (2,0.75);
   \vertex (c) at (2,-0.75);
   \vertex (2) at (3,2) {$\ell_{\alpha}$};
   \vertex (3) at (3,-2) {$\overline{\ell_{\beta}}$};
   \vertex (a2) at (6.5,0);
   \vertex (b2) at (8,0); 
   \vertex (12) at (5,0) {$Z'$};
   \vertex (22) at (9.5,0) {$Z$};
   \diagram*
   {
   (1) -- [boson, momentum=$p$] (a),
   (a) -- [fermion, edge label=$\nu_i$] (b),
   (a) -- [anti fermion, edge label'=$\nu_j$] (c),
   (b) -- [boson, edge label = $W$] (c),
   (b) -- [fermion] (2),
   (c) -- [anti fermion] (3),
   (12) -- [boson, momentum=\(p\)] (a2),
   (a2) -- [half left, fermion, edge label = $\nu_i$] (b2) -- [half left, fermion, edge label = $\nu_j$] (a2),
   (b2) -- [boson, momentum=\(p\)] (22),
   };
   \end{feynman}
 \end{tikzpicture}
 \caption{One-loop diagrams contributing to the coupling of 
 the $Z'$ boson to charged leptons (\textit{left}) and to kinetic and mass mixing between the $Z'$ and $Z$ bosons (\textit{right}).}
 \label{Fig:ZloopW}
 \end{figure}
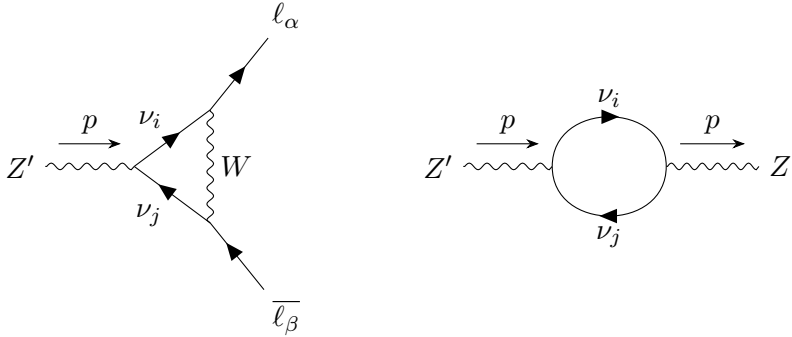

Since the neutrino mass eigenstates have components that couple both to the $Z$ and the $Z'$, mixing between the two gauge bosons will be induced at loop level~\cite{Holdom:1985ag} through the second diagram in Fig.~\ref{Fig:ZloopW}. 
The kinetic and mass mixings are described by the effective Lagrangian
\begin{align}
\mathcal{L}_{Z'Z} = -\frac{\sin{\epsilon}}{2}Z'_{\mu\nu}Z^{\mu\nu}+\delta m^2Z'_{\mu}Z^{\mu}\,.
\label{eq:Zmix}
\end{align}
Notice that these two terms
could be present already at the Lagrangian level 
after gauge symmetry breaking.
These would represent additional free parameters of the Lagrangian. However, these parameters do not contribute to the neutrino portal of interest here. Conversely, the neutrino mixing required for the neutrino portal does induce the $Z$-$Z'$ mixing at the loop level. Barring fine-tuned cancellations between the allowed free parameters at the Lagrangian level and the loop-induced contributions from neutrino mixing, the minimum contribution present in our set-up will be the latter. We will therefore set the tree-level parameters to zero and require that the loop-induced contributions are below the present experimental constraints on $Z$-$Z'$ mixing.
We find the following results for the mixing parameters: 
\begin{align}
\delta m^2 &= \frac{2}{(4\pi)^2}g'\frac{g}{\cos{\theta_W}}|U_{s4}|^2\left(1-|U_{s4}|^2\right)m_4^2\, f_{1}\,, \\
\sin{\epsilon} &= \frac{2}{(4\pi)^2}g'\frac{g}{\cos{\theta_W}}|U_{s4}|^2\left(1-|U_{s4}|^2\right)f_{2}\,, 
\label{eq:sineps}
\end{align}
where $f_1$ and $f_2$ are functions of $x\equiv m_4^2/p^2$, namely,
\begin{align}
f_{1}(x) &= \frac{1}{12}\bigg\lbrace 4x^2\left(1-x^{-1}\right)^3\coth^{-1}{(1-2x)}+2x-x^{-1}\log\left({x}\right)\nonumber \\
&\phantom{{}={}\frac{1}{12}\bigg\lbrace} -2\sqrt{x\left(4-x^{-1}\right)^3}\arctan{\left(\left(4x-1\right)^{-1/2}\right)}\bigg\rbrace\,, \\
f_{2}(x) &= -\frac{x^2}{6}\bigg\lbrace 4\left(2x-3+x^{-2}\right)\coth^{-1}{(1-2x)}+4+x^{-2}\log{(x)} \nonumber \\
&\phantom{{}= -\frac{x^2}{6}\bigg\lbrace} -2\sqrt{x^{-1}(4-x^{-1})}\left(2+x^{-1}\right)\arctan{\left(\left(4x-1\right)^{-1/2}\right)}\bigg\rbrace\,.
\label{eq:fKM}
\end{align}
%
For the purposes of this work $p^2\sim m_{\chi}^2$, and thus, $f_1$ and $f_2$ will only depend on the ratio of the masses of the heavy neutrino and the DM particle.
%
Following Ref.~\cite{Babu:1997st}, we first diagonalise the kinetic term through a non-unitary transformation and then perform a rotation to diagonalise the mass term.
The mass eigenstates $Z_1$ and $Z_2$ have masses given by
\begin{equation}
m_{Z_{1,2}}^2 = \frac{\sec^2{\epsilon}}{2}\left( m_Z^2+m_{Z'}^2
-2 \delta m^2\sin{\epsilon} \mp \Delta\right),
\label{eq:MZonetwo}
\end{equation}
%
where
\begin{align}
\Delta &= \sign\left(m_{Z'}^2-m_Z^2\left(1-2\sin^2{\epsilon}\right)
-2 \delta m^2\sin{\epsilon}\right) \nonumber\\
&\phantom{{}={}} \times \sqrt{m_Z^4+m_{Z'}^4+4 \delta m^4
-4 \left(m_Z^2+m_{Z'}^2\right) \delta m^2 \sin{\epsilon}
-2 m_Z^2 m_{Z'}^2\left(1-2\sin^2{\epsilon}\right)}\,.
\end{align}
%
From Eq.~\eqref{eq:MZonetwo}, one can easily verify that in the limit of small mass and kinetic mixing, i.e., 
$\delta m^2 \rightarrow 0$ and $\sin{\epsilon}\rightarrow 0$, 
the masses $m_{Z_1}\rightarrow m_Z$ and $m_{Z_2} \rightarrow m_{Z'}$.
After the full diagonalisation, we can write the $Z$ and $Z'$ in terms of the mass eigenstates $Z_1$ and $Z_2$ as follows:
\begin{align}
Z_{\mu}&=\left(\cos{\xi}-\tan{\epsilon} \sin{\xi}\right)Z_{1\mu}
-\left(\sin{\xi}+\tan{\epsilon}\cos{\xi}\right)Z_{2\mu}\,,\\
Z'_{\mu}&=\sec{\epsilon}\left(\sin{\xi}\, Z_{1\mu}+\cos{\xi}\, Z_{2\mu}\right),
\label{eq:ZstatesinZnewstates}
\end{align}
%
where $\xi$ is the angle related to the mass diagonalisation, 
which is defined through
\begin{align}
\tan{\left(2\xi\right)} = \frac{2\cos{\epsilon}
\left(m_Z^2 \sin{\epsilon}-\delta m^2\right)}
{m_{Z'}^2-m_Z^2\left(1-2\sin^2{\epsilon}\right)-2\delta m^2\sin{\epsilon}}\,.
\label{eq:Massmixangle}
\end{align}
%
The two angles $\xi$ and $\epsilon$ will control the phenomenology associated to the $Z$-$Z'$ mixing and consequently, the possible $Z'$ couplings 
to fermions.

The loop-induced kinetic mixing parameter $\eps$ depends solely on the ratio $x \approx m_4^2/m_\chi^2$, providing the coupling $g'$ and the element $U_{s4}$ of the neutrino mixing matrix are fixed (see Eqs.~\eqref{eq:sineps} and \eqref{eq:fKM}),
and increases with it.
Fixing $|\th_e| = 0.031$ and $\th_{\mu,\tau} = 0$, 
we find that for $x = 4$, which is the lowest value preventing
the $\chi\overline{\chi} \to \nu_i\overline{\nu_4}$, $i=1,2,3$, channels, and $g' = 1~(4\pi)$,
the mixing parameter $|\sin\eps|$ 
is of order of $10^{-6}~(10^{-5})$.
For values of $x$ as large as $10^4$  and $g' = 1~(4\pi)$,  the value of $|\sin\eps|$ does not exceed approximately $10^{-5}~(10^{-4})$. 

Generally, these values can be probed in beam dump and fixed target experiments searching for visible decay products (electrons and muons) of the $Z_2$ boson with mass between approximately 1~MeV and 1~GeV (see, e.g., \cite{Harnik:2012ni,Bauer:2018onh}). However, in the considered model the $Z_2$ decays mostly invisibly, either to a pair of the SM neutrinos or, if it is heavy enough, to a pair of DM particles, while its decays to charged leptons are suppressed. Thus, the bounds from fixed target experiments will not apply in this case. The supernova constraints cover nearly the same $Z_2$ masses, but a different range of $\eps \sim 10^{-10} - 10^{-7}$ \cite{Harnik:2012ni}, which thus are also avoided. 
For larger $Z_2$ masses, up to 100~GeV, collider experiments 
place the best constraints on $\eps \sim 10^{-4} - 10^{-3}$ 
(see, e.g., Ref.~\cite{Bauer:2018onh}).  
There exist also collider searchers for $Z_2$ decaying invisibly, 
which constrain $\eps \lesssim 10^{-3}$ for $m_{Z_2} < 8$~GeV~\cite{Lees:2017lec}.
These collider constraints are
above the values of the loop-induced
kinetic mixing parameter in our model. 
Finally, the much weaker constraint from the invisible $Z_1$ width, 
$\eps \lesssim 0.03$~\cite{Hook:2010tw},  is also evaded.

\begin{figure}
\centering
\includegraphics[width=7.6cm]{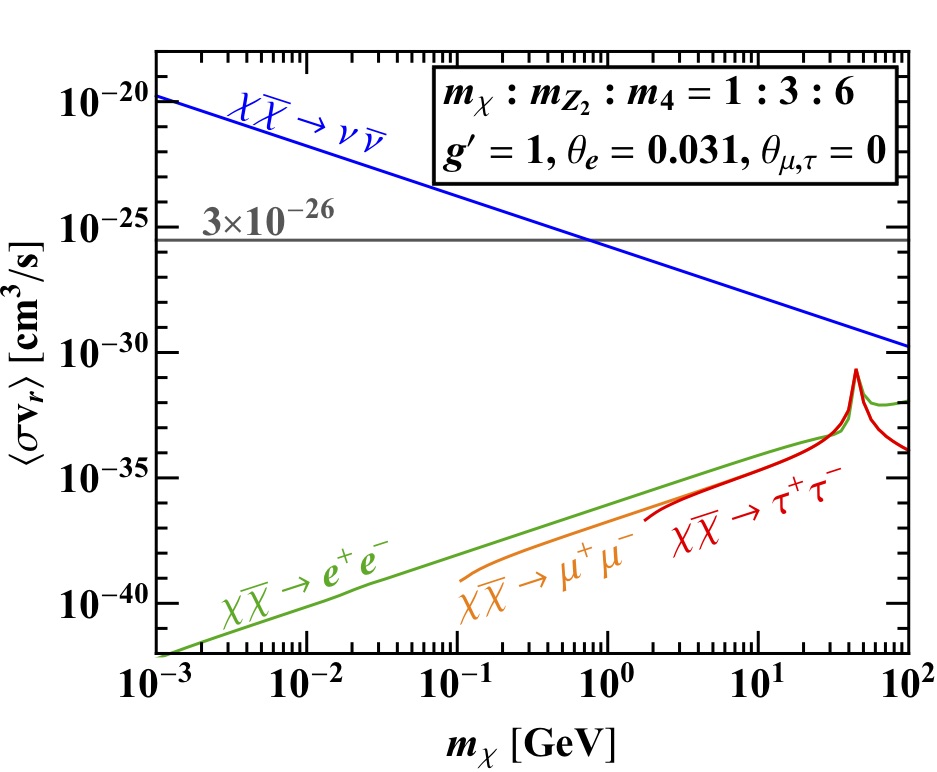}
\hspace{0.1cm}
\includegraphics[width=7.6cm]{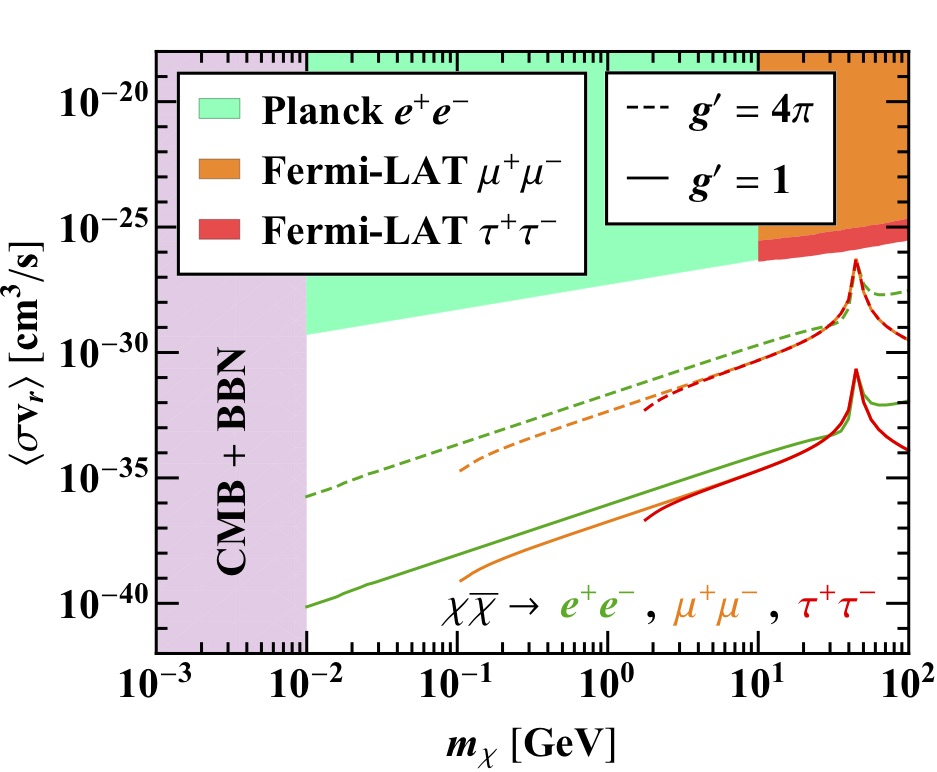}
\caption{Thermally averaged annihilation cross section multiplied by the relative velocity 
for DM annihilation into $e^+e^-$, $\mu^+\mu^-$, and $\tau^+\tau^-$. 
We have fixed $m_\chi : m_{Z_2} : m_4 = 1:3:6$, 
$g' = 1$, $\th_e = 0.031$, and $\th_{\mu,\tau} = 0$. 
The \textit{left} panel provides comparison with $\ev{\s\vr}$ for DM annihilation into neutrinos assuming the same set of model parameters. 
The \textit{right} panel displays the indirect detection constraints 
coming from Planck and Fermi-LAT. 
The lower bound $m_\chi \gtrsim 10$~MeV 
is set by observations of the CMB and BBN. See text for further details.}
\label{fig:mDMsigvEmiXgp1V}
\end{figure}

Together with the first diagram in Fig.~\ref{Fig:ZloopW}, the size of $\xi$ and $\epsilon$ will determine how relevant the DM annihilation to a pair of charged leptons is. We find that the tree-level annihilation to neutrinos dominates over that to charged leptons. In Fig.~\ref{fig:mDMsigvEmiXgp1V}, we show a particular example of this behaviour for  
$m_4 = 2m_{Z_2}$, $m_{Z_2} = 3m_\chi$, $g'=1$,
$|\th_e| = 0.031$, and $\th_\mu = \th_\tau = 0$. 
It is clear from this figure that the annihilation to charged leptons is unconstrained by current experimental searches. 
Note that the Planck and Fermi-LAT constraints 
shown in the right panel of Fig.~\ref{fig:mDMsigvEmiXgp1V}
assume a $100\%$ annihilation rate 
into a single SM channel.
%

\subsection{Results}
\label{sec:VectorResults}
\begin{figure}
\vspace{-2cm}
\centering
\includegraphics[width=7.5cm]{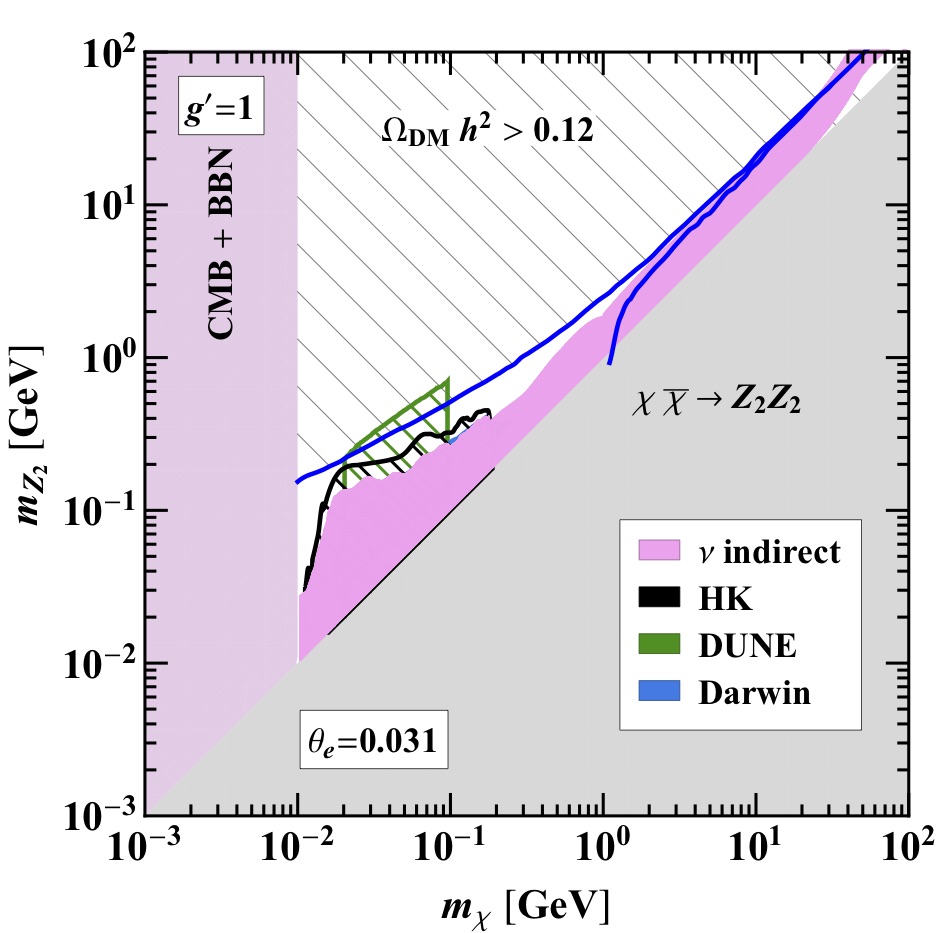}
\hspace{0.1cm}
\includegraphics[width=7.5cm]{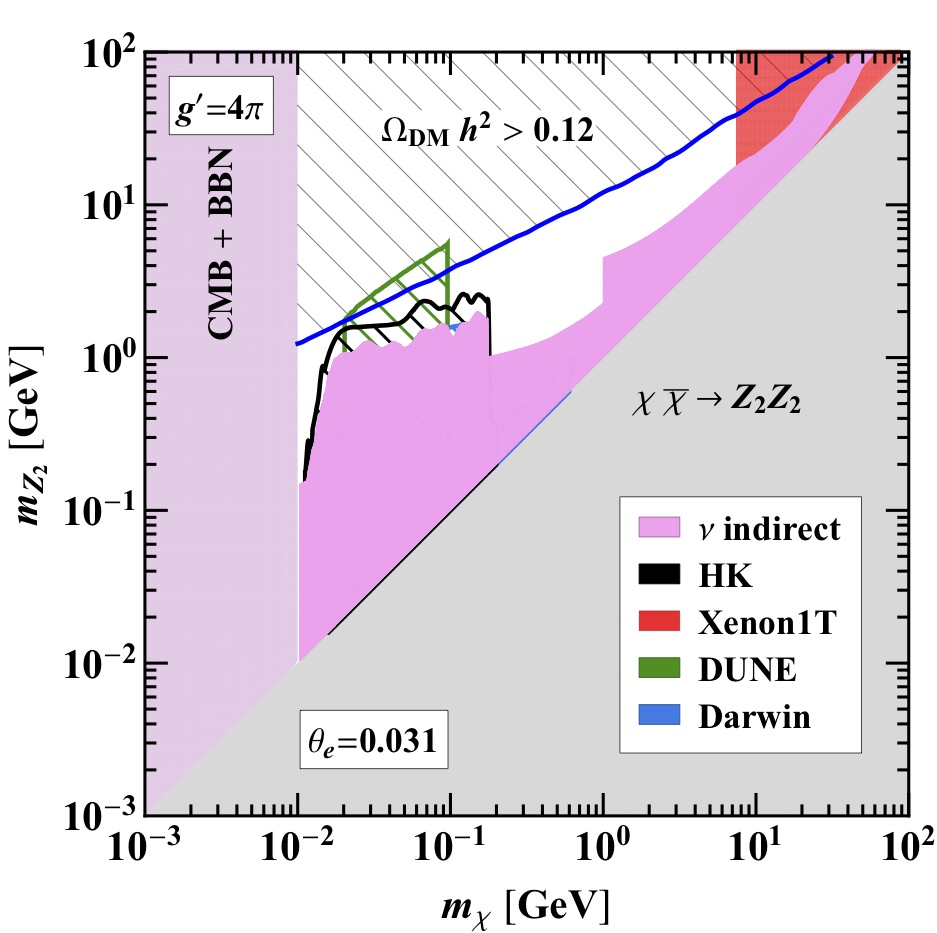} 
\includegraphics[width=7.5cm]{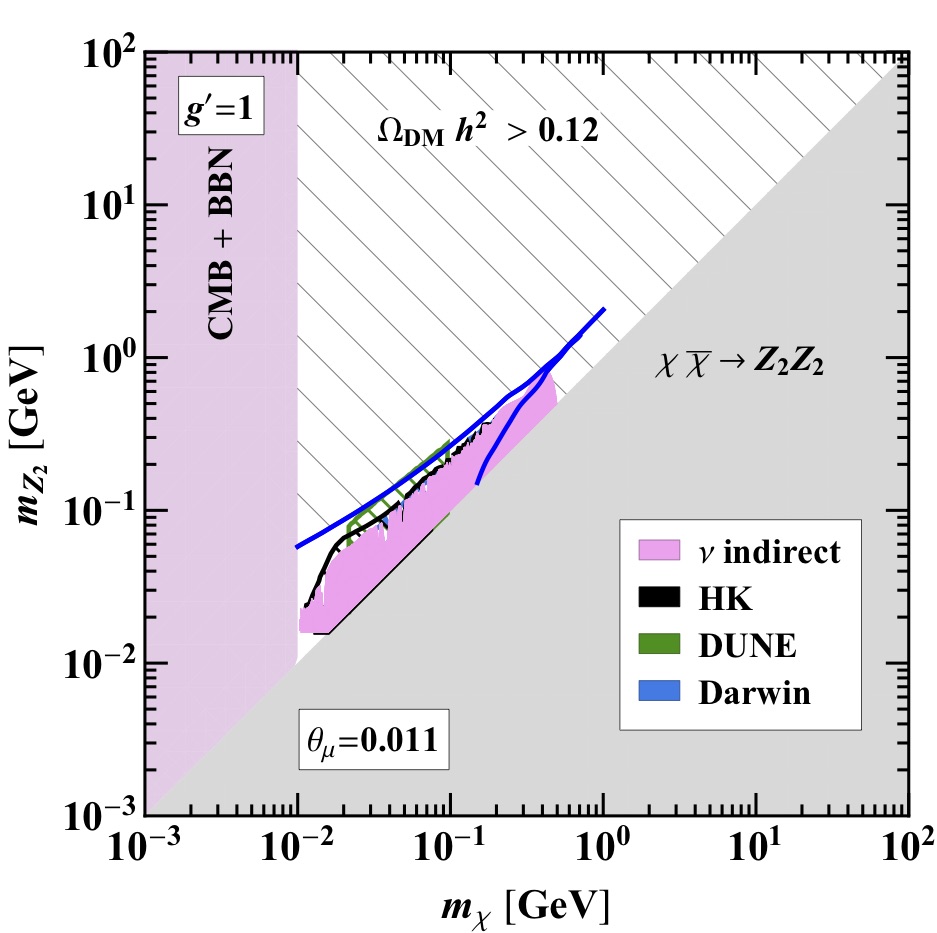}
\hspace{0.1cm}
\includegraphics[width=7.5cm]{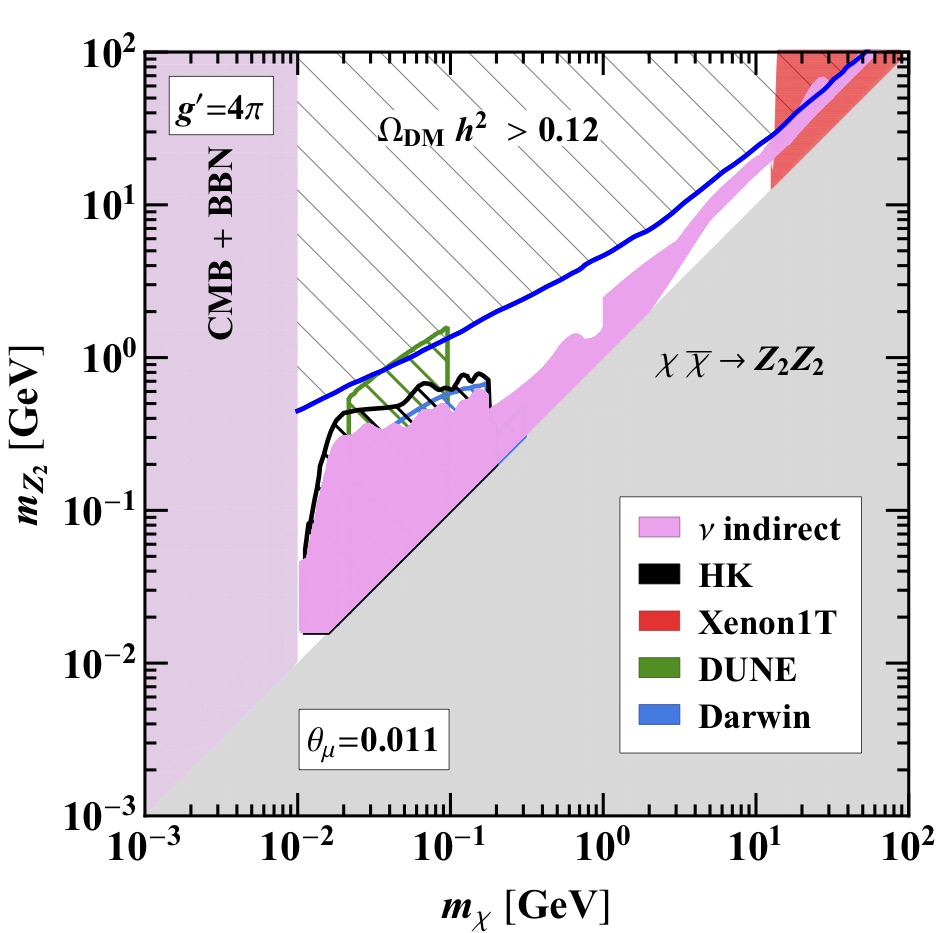} 
\includegraphics[width=7.5cm]{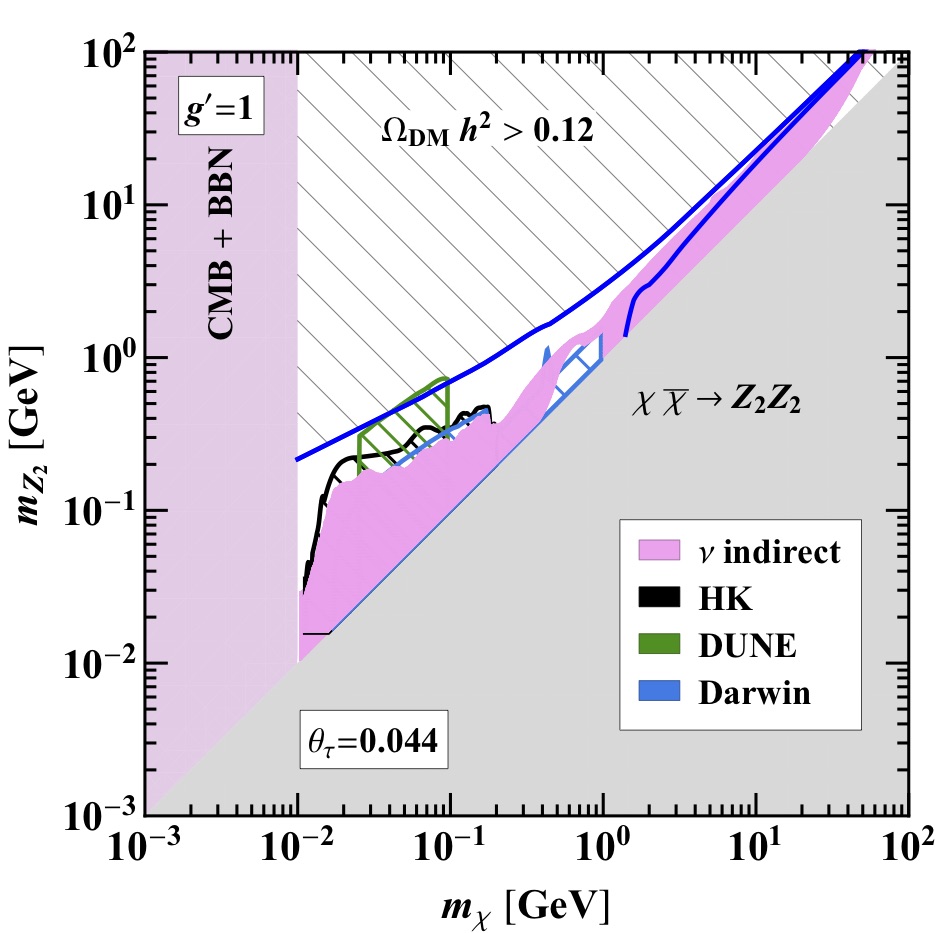}
\hspace{0.1cm}
\includegraphics[width=7.5cm]{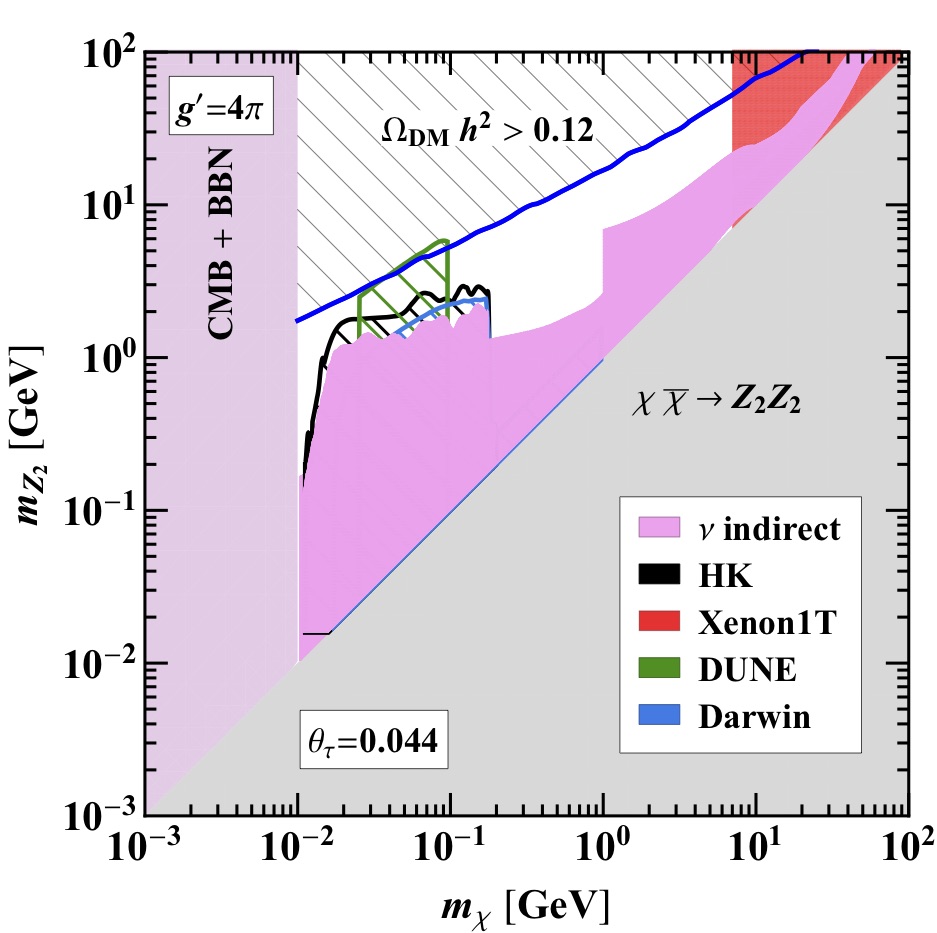} 
\caption{Constraints on the DM mass $m_\chi$ and $m_{Z_2}$. Along the blue lines, computed with \texttt{micrOMEGAs}, 
the DM relic density matches the observed value.
The coloured shaded regions are excluded by different experiments.
The lower bound $m_\chi \gtrsim 10$~MeV 
is set by observations of the CMB and BBN. See text for further details.}
\label{fig:mDMVector}
\end{figure}

The allowed regions of the parameter space in the $m_\chi$-$m_{Z_2}$ plane that satisfy cosmological, indirect and direct detection constraints for this model are presented in Fig.~\ref{fig:mDMVector} for $g'=1$ and $4\pi$, setting $\theta_\a \neq 0$ one at a time and keeping two other mixing angles fixed to zero. 
%
For definiteness, in the figure we set $m_4 = 2 m_{Z_2}$. Notice that this choice is not relevant for the interaction between the SM neutrinos and DM and only plays a role in the loop-induced processes that are sub-dominant. Nevertheless, if the $Z_2$ originates from a new $U(1)'$ gauge group, 
its mass $m_{Z_2}$, as well as that of the Dirac neutrino $m_4$, are generated
after the breaking of the symmetry. Thus, 
the natural expectation is that $m_4$ is not much heavier than $m_{Z_2}$ as long as the new gauge coupling $g'$ is $\mathcal{O}(1)$. 
Hence, unlike for the scalar example, it is not appropriate to set $m_4$ to a value above the electroweak scale 
while exploring (sub-)GeV $Z_2$ boson masses.

Below the electroweak scale constraints 
on the neutrino mixing parameters $\th_\a$ are \textit{a priori} 
much more stringent \cite{Atre:2009rg}. 
However, in the model under investigation 
the heavy neutrino decays mostly invisibly to 
either a SM neutrino and the $Z_2$ (if $m_4 > m_{Z_2}$),
or a SM neutrino and a pair of the DM particles (if $m_4 < m_{Z_2}$), 
assuming $g' \gtrsim 1$. This implies that the existing collider 
and beam dump constraints%
\footnote{If the heavy neutrino decays 
before reaching the detector, 
the constraints from beam dump experiments 
will not apply at all.}
should be rescaled 
with the corresponding branching ratios
and become even weaker than the non-unitarity constraints imposed previously for the scalar realisation. 
The bounds from peak searches in leptonic decays of pions and kaons will however apply, since they rely entirely on the kinematics of a two-body decay. 
Thus, the non-unitarity constraints actually dominate down to $m_4 \approx m_K \approx 0.5$~GeV, where $m_K$ is the kaon mass. 
In the region $m_4 \sim 0.01 - 0.4$~GeV, the bounds on $U_{e4}$ and $U_{\mu4}$
from peak searches are very stringent. We do not display them explicitly in Fig.~\ref{fig:mDMVector}, because 
they are $m_4$-dependent, while all the constraints shown in the figures have an extremely sub-leading dependence on $m_4$, as outlined above. Thus, Fig.~\ref{fig:mDMVector} is to be interpreted as generally valid for any neutrino mass $m_4 > m_K$. 

The blue line was calculated with \texttt{micrOMEGAs} and represents the DM and vector boson masses that will produce the correct relic abundance in a thermal scenario, while the masses in the upper hatched area would generate too much DM. A key difference with respect to the previous model is that here the DM annihilation cross section to neutrinos proceeds via an $s$-channel and thus is enhanced for $m_{Z_2} \sim 2 m_\chi$, as can be seen from Eq.~\eqref{eq:AnnCSVector}. This explains the second 
branch of the
blue line below the resonant condition in the panels with $g'=1$. A line where the relic abundance can be obtained below $m_{Z_2} = 2 m_\chi$ also occurs for $g'=4\pi$ but, since the cross section is larger, the relic abundance is achieved  for $m_{\chi} > 100$ GeV, which is ruled out by XENON1T. This resonant effect also explains the shape of the indirect detection constraints which follow the same trend.

Similar to the previous model in Section~\ref{sec:Scalar}, the direct detection constraints from XENON1T become relevant at large DM masses for $g'=4\pi$. However, even for values of the gauge coupling this large, we have checked that direct detection constraints from the elastic DM scattering off electrons are negligible. 

The complementarity between cosmological observables, DM, and neutrino experiments allows us to set very strong bounds on the DM and $Z_{2}$ masses for this particular realisation, ruling out significant portions of the parameter space. There are still allowed regions for larger values of the gauge coupling consistent with a thermal DM candidate that yields the observed DM relic abundance. However, future neutrino experiments such as DUNE will be able to probe down to the value for which the correct relic abundance is obtained in some parts of the parameter space.

\begin{figure}
\centering
\includegraphics[width=10cm]{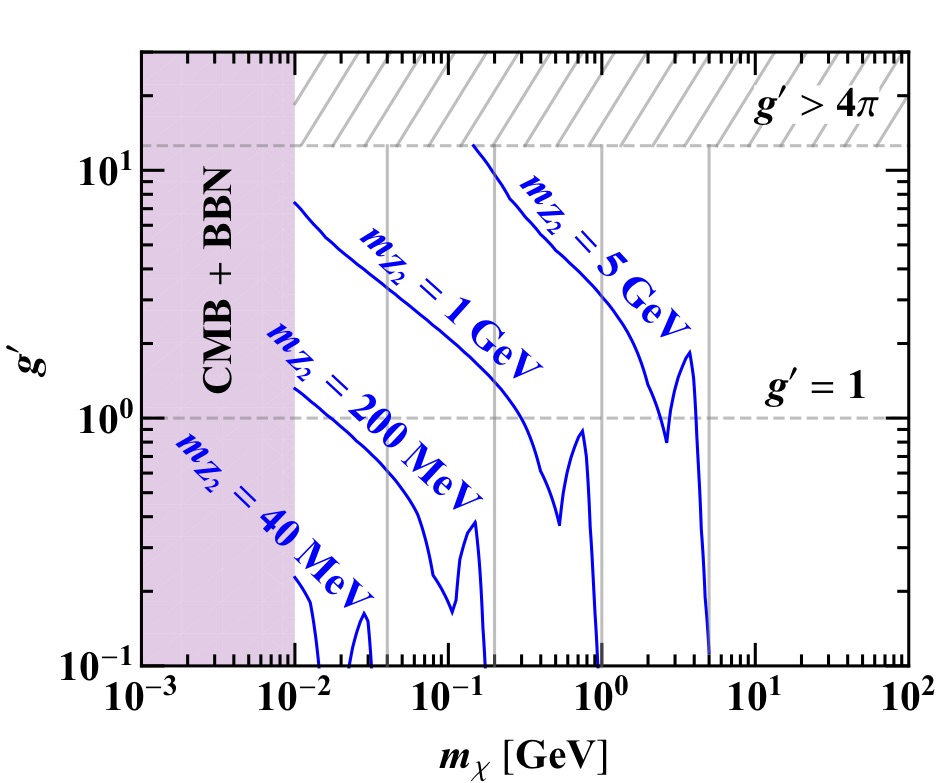}
\caption{Values of the DM mass $m_\chi$ and the coupling $g'$ required to reproduce the
observed relic abundance.
We have fixed $m_{Z_2} = 0.04$, $0.2$, $1$, and $5$~GeV, 
and have considered the representative case of  $\th_e = 0.031$, while keeping $\th_{\mu,\tau} = 0$.
Along (above) the blue lines
the DM relic density matches (is less than) the observed value. We do not consider $m_{\chi} > m_{Z_2}$ to ensure the neutrino portal regime. 
The lower bound $m_\chi \gtrsim 10$~MeV 
is set by observations of the CMB and BBN.}
\label{fig:mDMgpeMiX}
\end{figure}
%
It is worth noticing that the sensitivity of present and future neutrino detectors to DM annihilations into neutrinos is largely independent of the flavour to which the sterile neutrino dominantly couples. Indeed, regardless of the original flavour composition produced by the DM annihilations, neutrino oscillations will tend to populate all flavours with similar fractions when the flux arrives to the detector. The main differences between the three rows in Fig.~\ref{fig:mDMVector} are due to the different magnitude of the mixing allowed to the different flavours, with more stringent constraints applying for the mixing with muon neutrinos. 

Finally, in Fig.~\ref{fig:mDMgpeMiX}, we fix $m_{Z_2}$ 
to several values, namely, 
$m_{Z_2} = 0.04$, $0.2$, $1$, and $5$~GeV, and 
show the lines corresponding to the correct relic abundance 
in the $m_\chi$-$g'$ plane. 
These results were obtained using \texttt{micrOMEGAs}.
Small values of $g'$ are ruled out 
except for DM masses in the proximity of the resonance, 
i.e., when $m_{\chi} \approx m_{Z_2}/2$.
As can be seen from this figure, a lighter dark vector boson allows for 
smaller values of $g'$. 
For $m_{Z_2} \gtrsim 1$~GeV, values of $g' \gtrsim 1$ are required to yield the observed relic density, except for the resonance region. 
The dip towards $m_\chi \approx m_{Z_2}$ corresponds to 
opening of new DM annihilation channels at tree level.

\section{Conclusions}
\label{sec:conclusions}
Despite the tremendous improvement over the last years in the sensitivity of direct, indirect and collider searches for dark matter, its discovery still eludes us. An interesting possibility is that its interactions with SM particles happen dominantly with the neutrino sector. This option would not only explain our failure to detect any DM interactions (except gravitational) so far, it would also connect our two present experimental signals of physics beyond the SM. Indeed, a rich phenomenology that would stem from the connection of these two sectors has been explored and discussed in the literature. $SU(2)$ gauge invariance would naively dictate that neutrinos share all their interactions with their charged lepton counterparts, which are much easier to detect. We have therefore explored whether a dominant neutrino-DM interaction is allowed in simple gauge-invariant models without conflicting with searches through charged leptons.

We first explored the simplest scenario, in which DM couples to the full lepton doublet. We verified that, as long as the DM is heavier than the charged lepton(s) it couples to, the bounds from DM annihilation to charged leptons preclude DM-neutrino couplings sizeable enough to be probed, even ruling out all of the parameter space that would not lead to overclosure of the Universe. 
Alternatively, if DM couples to $\tau$ ($\mu$) and is lighter than the charged lepton, its phenomenology is dominated by the interaction with neutrinos. 
This region is constrained by present neutrino detectors and will be fully probed for certain DM masses by future experiments.

We have then explored the option of the neutrino portal to DM and showed, as an example, two specific realisations with scalar and vector couplings, respectively. In the neutrino portal DM couples directly to new heavy neutrinos. Indeed, their singlet nature makes them natural candidates to probe the dark sector since they are allowed to interact with it via relevant or marginal operators. These right-handed neutrinos are also a natural addition to the SM particle content so as to account for the evidence for neutrino masses and mixings. The mixing between the SM neutrinos and the new singlets will induce DM-neutrino interactions at tree level, but DM-charged lepton couplings only at loop level.

In the two realisations explored we find that it is indeed possible for neutrino detectors to place the most stringent and competitive bounds through searches for DM annihilations to neutrinos. Present searches at Super-Kamiokande, Fr\'ejus, or Borexino are ruling out large areas of the parameter space. Interestingly, future projects such as Hyper-Kamiokande, MEMPHYS, DARWIN, or DUNE will be able to probe the cross section very close and beyond the value required to explain the DM abundance solely by annihilation to SM neutrinos. These new searches will effectively cover most of the parameter space, probing if the right-handed singlet fermions that can explain the origin of neutrino masses also represent our best window to the discovery of the dark matter sector.

\section*{Acknowledgements}
We warmly thank B.~Zaldivar for extremely useful discussions. 
We would also like to acknowledge discussions with 
M.~Bauer, M.~Chala, M.~Hostert, A.~Plascencia, 
A.~Vincent, and S.~Witte. 
This work made extensive use of the HPC-Hydra cluster at IFT.
This work is supported in part by the European Union's Horizon 2020 research and innovation programme under the Marie Sklodowska-Curie grant agreements 674896-Elusives, 690575-InvisiblesPlus, and 777419-ESSnuSB, as well as by the COST Action CA15139 EuroNuNet. 
MB, EFM, and SR acknowledge support from the ``Spanish Agencia Estatal de Investigaci\'on'' (AEI) and the EU ``Fondo Europeo de Desarrollo Regional'' (FEDER) through the project FPA2016-78645-P; and the Spanish MINECO through the ``Ram\'on y Cajal'' programme and through the Centro de Excelencia Severo Ochoa Program under grant SEV-2016-0597. MB also acknowledges support from the G\"oran Gustafsson foundation. SP and AOD are also (partially) supported  by  the  European  Research  Council under ERC Grant “NuMass” (FP7-IDEAS-ERC ERC-CG 617143). SP would like to acknowledge partial support from the Wolfson Foundation and the Royal Society. SP and AOD would also like to thank the Instituto de F\'isica Te\'orica for kind hospitality during the completion of this work.

\bibliography{NDMportals_v2}

\end{document}